\documentclass[aps,superscriptaddress,showpacs,floatfix,nofootinbib]{revtex4-1}

\newcommand{\pPb}{p+{\rm Pb}\,}

\newcommand{\dAu}{d+{\rm Au}\,}
\newcommand{\hAu}{$^3${\rm He}+{\rm Au}\,}
\newcommand{\CuCu}{{\rm Cu}+{\rm Cu}\,}
\newcommand{\AuAu}{{\rm Au}+{\rm Au}\,}
\newcommand{\PbPb}{{\rm Pb}+{\rm Pb}\,}

\usepackage{hyperref}
\usepackage{color}
\usepackage{graphicx}	% Include figure filed
\graphicspath{%
  {./},%
}

\usepackage{xspace}	% Include xspace

\begin{document}

\title{Collective flow without hydrodynamics: simulation results for relativistic ion collisions}

\author{P. Romatschke} \affiliation{University of Colorado at Boulder}

\date{\today}

\begin{abstract}
Flow signatures in experimental data from relativistic ion collisions are usually interpreted as a fingerprint of the presence of a hydrodynamic phase during the evolution of these systems. In this work, flow signatures arising from event-by-event viscous hydrodynamics are compared to those arising from event-by-event non-interacting particle dynamics (free-streaming), both followed by a late-stage hadronic cascade, in \dAu, \hAu at $\sqrt{s}=200$ GeV and \pPb collisions at $\sqrt{s}=5$ TeV, respectively. For comparison, also \PbPb collisions at $\sqrt{s}=2.76$ TeV are simulated. It is found that non-hydrodynamic evolution can give rise to equal or larger radial flow than hydrodynamics with $\eta/s=0.08$ in all simulated collision systems. In light-on-heavy-ion collisions, free-streaming gives rise to triangular and quadrupolar flow comparable to or larger than that from hydrodynamics, but it generally leads to considerably smaller elliptic flow.  As expected, free-streaming leads to considerably less elliptic, triangular and quadrupolar flow than hydrodynamics in nucleus-nucleus collisions, such as event-by-event \PbPb collisions at $\sqrt{s}=2.76$ TeV.
\end{abstract}

\maketitle

\section{Introduction}

One of the recent great successes of high-energy nuclear physics has been the understanding that the hot QCD matter created in high-energy nucleus-nucleus collisions behaves like a strongly-coupled, low-viscosity fluid rather than a weakly-coupled, almost ideal gas of quarks and gluons. This finding rests on the fact that several, independent observational probes of the hot QCD matter, such as for example the strong collective flow and strong jet quenching, can all naturally be explained theoretically under the single assumption of a hydrodynamic phase early in the evolution of the hot QCD matter. 

However, the sizable flow signals that have been measured in systems created in light-on-heavy-ion collisions at RHIC and the LHC \cite{Abelev:2012ola, Aad:2012gla, Adare:2013piz, Chatrchyan:2013nka, Shengli,Khachatryan:2015waa} have come as a surprise to many experts in the field. There are currently two competing interpretation for this finding. On the one hand, the flow signals measured in light-on-heavy-ion collisions may just have the same origin than those in nucleus-nucleus collisions, namely the presence of a hydrodynamic phase (see Refs.~\cite{Bozek:2011if,Nagle:2013lja,Schenke:2014zha,Schenke:2014gaa,Kozlov:2014fqa,Romatschke:2015gxa,Bozek:2015qpa,Yan:2015fva} for work along those lines). Because the systems created in light-on-heavy-ion collisions are very small and short-lived compared to those created in nucleus-nucleus collisions, one may, however, entertain a different hypothesis, namely that flow is being created by some other, non-hydrodynamic mechanism (see Refs.~\cite{Gyulassy:2014cfa,Dumitru:2014yza,He:2015hfa,Bozek:2015swa,Zhou:2015iba} for work along those lines). 

In view of these competing interpretations, the present work tries to answer the following question: How different would flow observables be if the systems created in relativistic ion collisions \emph{never} go through a hydrodynamic phase, but still experience expansion and cooling in the hot phase and interactions in the low-temperature, hadron gas phase? 

To answer this question, the dynamics of the hot QCD matter phase is alternatively described by two extreme (classical) opposites: hydrodynamics and a non-interacting gas of free-streaming particles. By keeping all steps of the model simulations the same but only switching from a hydrodynamic to a free-streaming description (with the same equation of state) allows one to make fully transparent comparisons between hydrodynamic and non-hydrodynamic results. The extreme nature of the two opposites employed also guarantee that any intermediate case between strong and weak interactions must be bounded by the results found in this study.

It is important to stress the relation of the present study to previous works. For instance, Refs.~\cite{Gyulassy:2014cfa,Dumitru:2014yza} proposed an initial state effect originating in QCD to explain the observed elliptic flow in light-on-heavy-ion collisions, a mechanism that seems to be in tension with more recent experimental results for multi-particle correlations \cite{Khachatryan:2015waa}. This line of work is complementary to the present study, since in the present study the emphasis is on non-hydrodynamic flow generated during the hot QCD phase

Refs.~\cite{He:2015hfa,Bozek:2015swa} describe the hot QCD matter phase using the phenomenological, non-hydrodynamic AMPT model including strings, hard particles and effective interactions, as well as hydrodynamics. AMPT seems to match the experimental results from \pPb collisions at $\sqrt{s}=5.02$ TeV with a very small effective scattering cross-section, thus suggesting the possibility a non-hydrodynamic explanation for the observed flow signal. However, the fact that AMPT has many different phenomenological ingredients makes the interpretation of this result somewhat challenging. The main difference to Refs.~\cite{He:2015hfa,Bozek:2015swa} of the present study is the fully transparent nature of the non-hydrodynamic description employed here. 

Finally, Ref.~\cite{Zhou:2015iba} uses the hadronic cascade model URQMD to describe \pPb collisions at $\sqrt{s}=5.02$ TeV. It was found that purely hadronic interactions cannot describe experimental data in \pPb collisions, but no comparison with hydrodynamics using the same equation of state as in the URQMD model was attempted. The present study is similar to Ref.~\cite{Zhou:2015iba} in that a hadronic cascade code is employed to describe the low temperature system evolution. A key differences to Ref.~\cite{Zhou:2015iba} in the present study is that besides purely hadronic evolution (applicable to the low temperature phase), also two different scenarios for the hot QCD phase (hydrodynamics and free streaming) are considered.

The remainder of this work is organized as follows: In Section \ref{sec:meth}, the details for the model setup, initial conditions, equation of state, and calculation of final observables are given. Also, this section contains a warm-up example of comparing free-streaming and hydrodynamics in the case of collisions of smooth nuclei. In Section \ref{sec:res}, results from simulating event-by-event collisions of granular light nuclei on heavy nuclei (\pPb, \hAu and \dAu), as well as nucleus-nucleus collisions (\PbPb) are shown. The summary and conclusions are presented in Sec.\ref{sec:summ}, while the special case of an analytically solvable comparison between ideal fluid dynamics and free-streaming is discussed in appendix \ref{sec:app}.

\section{Methodology}
\label{sec:meth}

\subsection{Stage 1: Initial Conditions and Equation of State}
\label{sec:IS}

Initial conditions are prepared for the bulk energy-momentum tensor $T^{ab}$. For simplicity, only boost-invariant dynamics will be considered, which is most easily implemented by using an expanding space-time described through the Milne coordinates:
\begin{equation}
\tau=\sqrt{t^2-z^2}\,,\quad \xi=\frac{1}{2}\ln \frac{t+z}{t-z}\,,\quad
{\bf x_\perp}=\left(\begin{array}{c}x\\y\end{array}\right)\,.
\end{equation}
For a boost-invariant system, the dynamics is invariant under translations in $\xi$, so that all quantities only depend on $\tau,{\bf x_\perp}$; thus, initial conditions amount to specifying $T^{ab}(\tau=\tau_0,{\bf x_\perp})$ at some time $\tau_0$. 

Semi-realistic initial conditions for $T^{ab}$ have been calculated under certain assumptions of the collision dynamics \cite{Schenke:2012wb,Berges:2013fga,vanderSchee:2013pia,Gelis:2013rba,Kurkela:2014tea,Bantilan:2014sra,Chesler:2015wra}. The collection of these amount to a large class of (generally non-equilibrium) initial conditions that could be implemented for $T^{ab}(\tau=\tau_0)$. The goal of this study, however, is the comparison between the subsequent evolution of $T^{ab}(\tau,{\bf x_\perp})$, and thus arguably the simplest possible initial condition will be implemented in the following: the case of thermal equilibrium with zero local flow in the transverse ${\bf x_\perp}$ plane. In this case, the initial conditions are fully specified by the energy density $T^{00}({\bf x_\perp})$ and the equation of state (e.g. through the functional relation of the pressure $P(\epsilon)$ to the energy density $\epsilon$). It should be stressed that an equilibrium initial condition (or more precisely an initial condition with zero momentum anisotropy) is not required for the comparison. Also, it should be pointed out that a condition with zero momentum anisotropy is generated 'by accident' at some point during the time evolution for initially prolate momentum distributions (cf.~\cite{Strickland:2014pga}). In this sense, the initial conditions chosen above do not actually require equilibration of the system to happen before $\tau=\tau_0$, but correspond to a whole class of non-equilibrium initial conditions at specially chosen instances in time.

\begin{figure}[t]
\includegraphics[width=0.7\linewidth]{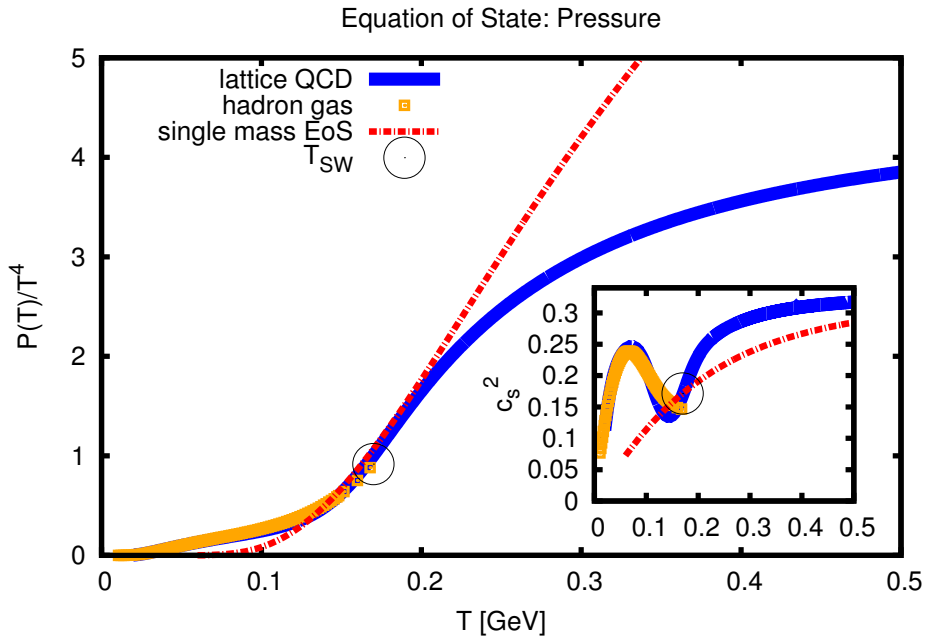}
\caption{\label{fig:one} Pressure versus temperature for different equations of state. Shown are results for the hadron resonance gas and the one-component massive gas with parameters adjusted to match the value and slope of the hadron gas pressure at $T=T_{SW}$. Inset shows dependence of speed of sound square $c_s^2$. For illustrative purposes, the full lattice QCD result (from Ref.~\cite{Borsanyi:2013bia}) is also shown. }
\end{figure}

Within hydrodynamics, arbitrary equations of state are easily implemented. In the case of classical particle dynamics described through a Boltzmann equation, this is considerably harder to do (see e.g. Ref.~\cite{Romatschke:2011qp} on how to implement generic equations of state through generalized Boltzmann equations). However, since the present study aims at mapping the hot QCD dynamics onto a hadron gas cascade evolution at low temperatures, the absolute minimum requirement in order to conserve energy and momentum throughout the whole evolution is to implement an equation of state that smoothly matches onto the equation of state from a hadron gas at some predefined switching temperature $T=T_{SW}$. This can be achieved by considering the equation of state generated by a number of $Z$ particle with mass $m$:
\begin{eqnarray}
\label{eq:eos}
\epsilon = \frac{Z\,  m^2 T}{2\pi^2}\left(3 T K_2\left(\frac{m}{T}\right)+m 
K_1\left(\frac{m}{T}\right)\right)\,,
\quad
p = \frac{Z\,  m^2 T^2}{2\pi^2}K_2\left(\frac{m}{T}\right)\,.
\end{eqnarray}
Choosing to match at $T_{SW}=0.17$ GeV, it is found that the hadron gas pressure as well as its first derivative can be matched at $T=T_{SW}$ by choosing $m=0.779$ GeV and $Z=116$. Figure \ref{fig:one} shows a comparison for the pressure from the massive gas, the pressure from the sum of all known hadron resonances up to masses of $2.2$ GeV in the Particle Data Book, and the pressure calculated in lattice QCD by the BMW collaboration \cite{Borsanyi:2013bia}. As can be seen, the fact that the value of the pressure as well as its first derivative match between the hadron gas equation of state and the one-component gas imply that a smooth transition from one description to the other is possible. For this reason, the one-component gas equation of state with parameters $m=0.779$ GeV and $Z=116$ will be adopted for the rest of this work, both in the non-interacting and hydrodynamic case.

One could worry that in the case of free-streaming the system would generically be far from equilibrium, and an equation of state for such as system could not be defined.  However, non-perturbative quantum field theory studies provide evidence that an equilibrium equation of state relating the energy density to the temperature is present after a very short time-scale even for otherwise out-of-equilibrium situations \cite{Berges:2004ce}. Thus, even far from equilibrium systems can reasonably be expected to possess a relation between energy density and temperature that is given by the equilibrium equation of state.

The initial conditions for the energy density are taken from two sources: first, a smooth optical Glauber model which is used to demonstrate key features of the free-streaming results in a simple setting. And second, from a more realistic event-by-event Monte-Carlo Glauber event generator taking into account the collision system composition and collision energy. 
In the first case, the different nuclei are modeled by employing an overlap function 
\begin{equation}
\label{eq:overlap}
T_A(x,y)=\int_{-\infty}^\infty dz \left[1+ e^{-(|{\bf x_\perp}|^2+z^2-R)/a}\right]\,,
\end{equation}
with $R,a$ the charge radius and skin depth parameters (in the following, $R=4.16$ fm and $a=0.606$ fm for $^{63}$Cu will be used). The initial condition for the energy density in this case is then given by
\begin{equation}
\label{pre-flowED}
\epsilon(\tau_0,{\bf x_\perp})=E_0 T_A(x+\frac{b}{2},y)T_A(x-\frac{b}{2},y)\,, 
\end{equation}
where $b$ controls the impact parameter of the collision and $E_0$ is an overall constant controlling the total multiplicity (entropy) simulated.

In the second case, initial conditions for the energy-density $\epsilon(\tau_0,{\bf x_\perp})$ for each event are constructed as follows. Using Woods-Saxon distribution functions for the heavy ions such as ${\rm Au,Pb}$ \cite{DeJager1974479,DeVries1987495}, the Hulthen wavefunction for the deuteron (cf.~\cite{Adare:2013nff}) and realistic calculations for the $^{3}{\rm He}$ wavefunction \cite{Carlson:1997qn}, probability distributions of the nucleons within the nuclei of interest (cf.~\cite{Nagle:2013lja}) are obtained. Using a Monte-Carlo Glauber \cite{Miller:2007ri}, these probability distributions are mapped to positions of individual nucleons in the transverse ($x,y$) plane on an event-by-event basis implementing a hard-core repulsive potential of radius $0.4$ fm between nucleons. The positions of nucleons undergoing at least one inelastic collisions are recorded (``participants'') and converted into a density function $R^2({\bf x})$ by assuming that each participant contributes equally as a Gaussian with a width of $w=0.4$ fm (to match the RMS radius of a single nucleon). The initial condition for the energy density is then assumed to be given as
\begin{equation}
\label{Eq:ed}
\epsilon(\tau_0,{\bf x})=E_0 R^2({\bf x})\,,
\end{equation}
with $E_0$ again an overall constant (dependent on $\tau_0$, collision energy and collision system) that is related to the total multiplicity of the event. Typically 100 initial conditions are generated for each collision system.

\subsection{Stage 2, Option a: Hydrodynamics}

Once the initial conditions for the energy-momentum tensor are specified, these can be converted into hydrodynamic degrees of freedom via the following decomposition of the energy-momentum tensor:
\begin{equation}
\label{eq:tabhy}
T^{ab}=\epsilon u^a u^b-(P-\Pi)\left(g^{ab}-u^a u^b\right)+\pi^{ab}\,,
\end{equation}
where $\epsilon,P$ are the local (equilibrium) energy density and pressure, $u^a$ is the local fluid four velocity, $g^{ab}$ is the metric tensor and $\Pi,\pi^{ab}$ are the shear and bulk stress tensors, respectively. Such a decomposition of the energy-momentum tensor is possible in most cases, with the exception of out-of-equilibrium quantum states (see e.g. Ref~\cite{Arnold:2014jva} for more discussions). For the case of equilibrium conditions at hand, the initial conditions imply $u^a=(1,0,0,0),\Pi=\pi^{ab}=0$ and $\epsilon(\tau_0,{\bf x_\perp})$ given by the initial conditions discussed in Sec.\ref{sec:IS}.

Once the initial conditions have been specified, the hydrodynamic equations of motion $\nabla_a T^{ab}=0$ have to be solved. To do this, one first needs to specify the constitutive relations that e.g. connect the shear tensors to gradients of the fundamental hydrodynamic degrees of freedom $\epsilon,u^a$. Fortunately, recent progress in relativistic fluid dynamics (which to a large extent has been fueled by access to strongly coupled field theory dynamics from the gauge/gravity duality conjecture) has led to a complete characterization of all possible terms that can appear to a certain order in gradients (see e.g. Refs.~\cite{Baier:2007ix,Bhattacharyya:2008jc,Romatschke:2009kr,Rangamani:2009xk}). 
Once the constitutive relations have been specified, one still needs an algorithm to actually solve the hydrodynamic equations of motion numerically. The standard approach in relativistic dissipative fluid dynamics, which has been fully developed in the past 10 years, is to use causal second-order fluid dynamics (see Refs.~\cite{Romatschke:2009im,Schafer:2009dj,Heinz:2013th} for reviews on this subject). 

Finally, there are transport coefficient functions appearing in the hydrodynamic equations of motion. At zeroth order in gradients (ideal fluid dynamics), the only such quantity is the speed of sound $c_s$, which is fully specified through the equation of state $c_s^2\equiv \frac{dP}{d\epsilon}$. At first order in gradients (Navier-Stokes fluid dynamics), there are the shear and bulk viscosity coefficients, denoted as $\eta,\zeta$, respectively. For simplicity, in this study only constant values for the ratio of shear viscosity over entropy density $s$ will be adopted, and the bulk viscosity will be set to zero. Finally, at second-order in gradients there are an additional $11$ transport coefficient in flat space-times \cite{Romatschke:2009kr}, and for simplicity most of these will again be set to zero, except for the relaxation time $\tau_\pi=\frac{4 \eta}{\epsilon+P}$.

With these specifications, the equations of motion are solved using the publicly available VH2+1 code package \cite{Luzum:2008cw,Romatschke:2015gxa}, version 2.0, on a two-dimensional space grid with lattice spacing of $\Delta x\sim 0.1$ fm. Every $0.25$ fm/c during the evolution, the local temperature of fluid cells is monitored and once a fluid cell cools below the switching temperature $T_{\rm SW}$, information about the cell's location as well as the value of $\epsilon,u^a,\Pi,\pi^{ab}$ is stored. The collection of all these cells' locations $(\tau,{\bf x_\perp})$ defines the switching hypersurface $\Sigma$, which will eventually be used to initialize the low-temperature hadron gas dynamics (see \ref{sec:s3}).

\subsection{Stage 2, Option b: Free-Streaming}

While the above hydrodynamic option to describe the bulk system dynamics is quite standard, this work proposes an ``option b'' for the dynamics: non-interacting free particle dynamics. In this case, the energy-momentum tensor is given in terms of the one-component on-shell particle distribution function $f(\tau,{\bf x_\perp},\xi,{\bf p_\perp},p^\xi)$ as 
\begin{equation}
\label{eq:tabkin}
T^{ab}=\int \frac{d^2 p_\perp d p^\xi \tau}{(2 \pi)^3} \frac{p^a p^b}{p^\tau} f(\tau,{\bf x_\perp},\xi,{\bf p_\perp},p^\xi)\,,
\end{equation}
where for on-shell massive particles $p^\tau=\sqrt{m^2+p_\perp^2+\tau^2 p^{\xi 2}}$. The distribution function will be taken to be a solution to the classical Boltzmann equation in the non-interacting (free-streaming limit) \cite{Romatschke:2011qp}:
\begin{equation}
\label{myBoltz}
p^a \partial_a f - \frac{2 p^\xi p^\tau}{\tau}\partial_\xi^{(p)}f = 0\,,
\end{equation}
where $\partial_a^{(p)}\equiv \frac{\partial}{\partial p^a}$. This equation is readily solved using the method of characteristics, finding the general solution
\begin{equation}
\label{eq:fsol}
f=f\left({\bf p_\perp},p_\xi,{\bf x_\perp}-\frac{\tau {\bf p_\perp} p^\tau}{p_\perp^2+m^2},\xi+\ln\left[\frac{p^\tau}{p_\xi}+\frac{1}{\tau}\right]\right)\,.
\end{equation}

Also, the implementation of equilibrium initial conditions is straightforward. Given an equation of state, the energy density defines a local equilibrium temperature $T=T(\epsilon)$ and an equilibrium solution for the particle distribution function for Eq.~(\ref{myBoltz}) can be shown to be given by
\begin{equation}
\label{eq:feq}
f_{\rm eq}=Z e^{-p^a u_a /T}\,,
\end{equation}
where $u^a$ is the local macroscopic (not necessarily fluid) four velocity with respect to some global laboratory frame, and $Z$ is the effective number of degrees of freedom first introduced in 
Eq.~(\ref{eq:eos}). It is straightforward to show that inserting Eq.~(\ref{eq:feq}) into Eq.~(\ref{eq:tabkin}) leads to the results given in Eq.~(\ref{eq:eos}). Evaluating Eq.~(\ref{eq:feq}) for $T=T\left(\epsilon({\bf x_\perp})\right)$ in the transverse plane then fully specifies the initial conditions for the free-streaming dynamics.

For the case of boost-invariant dynamics, and equilibrium initial conditions with $u^a(\tau_0)=(1,0,0,0)$ given at $\tau=\tau_0$, the solution to Eq.~(\ref{myBoltz}) at any later time may then be analytically written as\footnote{Note that the result correspond the form used in anisotropic hydrodynamics \cite{Strickland:2014pga}.}.
\begin{equation}
f(\tau,{\bf x_\perp},\xi,{\bf p_\perp},p^\xi)=Z \exp{\left[-p_0^\tau/T\left({\bf x_\perp}-\frac{{\bf p_\perp} (\tau p^\tau-\tau_0 p^\tau_0)}{p_\perp^2+m^2}\right)\right]}\,,\quad p^\tau_0=\sqrt{p_\perp^2+m^2+p_\xi^2/\tau^2_0}\,.
\end{equation}
From the solution at time $\tau$, one can evaluate the energy-momentum tensor $T^{ab}$ and from the energy-momentum tensor one can find the local energy density, flow velocity, shear and bulk stress tensors using the decomposition in Eq.~(\ref{eq:tabhy}). (Note that since the particle dynamics is classical, a decomposition along the lines of Eq.~(\ref{eq:tabhy}) is always possible even for far-from-equilibrium systems \cite{Arnold:2014jva}). Using the same routines as in the hydrodynamic framework, the local temperature is monitored and a switching hypersurface $\Sigma$ can again be defined as those space-time points which have $T=T_{\rm SW}$ (in practice, and for better comparability, the same two-dimensional lattice as in the hydrodynamic framework with $\Delta x\sim 0.1$ fm and a time-increment between steps of $0.25$ fm/c is used). The quantities $\epsilon,u^a,\Pi,\pi^{ab}$ are stored along the hypersurface and thus the final information available is exactly equal to that from the hydrodynamic framework.

\subsection{Stage 3: Kinetic Freeze-Out and Hadron Cascade}
\label{sec:s3}

Using information from the switching hyper-surface from either the hydrodynamic or free-streaming evolutions in the hot phase, the low temperature phase is simulated through a hadronic cascade code (B3D, \cite{Novak:2013bqa}). Using the hyper-surface information to boost to the rest frame of each cell, the cascade is initialized with particles in the rest frame drawn from a Boltzmann distribution at a temperature $T_{\rm SW}$ with modifications of the momentum distribution to include deformations from viscous (both shear and bulk) stress tensors (see \cite{Pratt:2010jt} for details). Specifically, for a particle with mass $\mu$, the distribution in the local rest frame is assumed to be of the form 
\begin{equation}
\label{fullmatching}
f({\bf p})=\exp{\left[-\sqrt{\mu^2+{\bf k^2}}/T^\prime\right]}, \quad
p_i=\left(\delta_{ij}+\lambda_{ij}\right)k_j\,,
\end{equation}
with $\lambda_{ij}$ controlling the size of the shear and/or bulk corrections to the stress-energy tensor. Note that for $\lambda_{ij}\ll1$, one finds $f({\bf p})-f_{\rm eq}({\bf p})\propto f_{\rm eq}({\bf p}) \frac{p_i p_j \lambda_{ij}}{p_0}$ \cite{Pratt:2010jt} with $T^\prime=T$. However, even for small values of the stress tensors, the distribution function does not recover 'quadratic ansatz' form \cite{Luzum:2008cw} , because of the additional power of $p_\mu u^\mu$ in the denominator resulting from expanding Eq.(\ref{fullmatching}). It should be stressed that this procedure ensures that the complete stress tensor $T^{ab}$ (not just its ideal fluid part) is matched across the hypersurface boundary. In particular, this implies that no assumption about equilibrium is made at the switching hypersurface: arbitrary deviation from equilibrium, parametrized through large dissipative tensor components, are allowed. If only the ideal fluid part of the stress-energy tensor is matched, this would lead to a 'fake' collective flow signal, such as a discontinuous elliptic flow component (cf.~\cite{Broniowski:2008qk}). See the discussion in Appendix \ref{app:two} for more details.

The cascade code B3D includes hadron resonances in the Particle Data Book up to masses of $2.2$ GeV, which interact via simple s-wave scattering with a constant cross-section of $10$ mb as well as scattering through resonances (modeled as a Breit-Wigner form). Once the resonances have stopped interacting, one can obtain final charged hadron multiplicities $\frac{dN_{\rm ch}}{d Y}$, mean charged particle momentum $\langle p_T\rangle$ and flow coefficients $v_n(p_T)$ for $n\geq 1$ from summing over individual particles with momenta ${\bf p}$. 
Specifically,
\begin{eqnarray}
&\frac{dN_{\rm ch}}{2\pi p_T dY dp_T}=\frac{\sum^{\rm ch.\ particles}_{\rm in\ p_T\ bin}}{2 \pi p_T \Delta_T \Delta_Y}\,,\quad
\frac{dN_{\rm ch}}{dY}=\int_0^\infty dp_T \frac{dN_{\rm ch}}{dY dp_T}\,,\quad
\langle p_T\rangle=\frac{\int_0^\infty dp_T p_T \frac{dN_{\rm ch}}{dY dp_T}}{\frac{dN_{\rm ch}}{dY}}
&\nonumber \\
&
|v_n|(p_T)=\sqrt{s_n(p_T)^2+c_n(p_T)^2}\,,\quad
\left(\begin{array}{cc}
s_n(p_T)\\
c_n(p_T)\end{array}\right)=
\frac{\sum^{\rm ch.\ particles}_{\rm in\ p_T\ bin}
\left(\begin{array}{cc}
\sin(n\phi))\\
\cos(n\phi)\end{array}\right)}
{\sum^{\rm ch.\ particles}_{\rm in\ p_T\ bin}}
\,,\quad
\phi\equiv \arctan\left(\frac{p_y}{p_x}\right)\,,&
\end{eqnarray}
where $\Delta_T=80$ MeV, $\Delta_Y=2$ are the width of bins for particle $p_T$ and rapidity $Y$, respectively. Note that since the cascade is applied to a boost-invariance case, the large $\Delta_Y$ value is of no significance. In practice, a sum over both particles and anti-particles and division of the spectra by two is performed, in order to increase statistics. For every hydrodynamic evolution event, at least 100,000 B3D events are run to increase statistics. In doing so, 
the sums in the definition of $v_n$ above are extended over all B3D events, thereby explicitly ignoring fluctuations arising from hadronic decays. After thus obtaining results for $\frac{dN_{\rm ch}}{2\pi p_T dY dp_T}$ and $v_n(p_T)$ for each hydrodynamic event, an event average to obtain the event-by-event mean and event-by-event fluctuation is performed. Results both for the case of hydrodynamic and free-streaming hot phase dynamics are reported on in the following. It should be noted that because B3D enforces detailed balance, there is no baryon annihilation simulated and as a consequence proton yields are too high. For this reason, the results reported for protons below should be interpreted with care.

\subsection{Warm-up: Collisions of idealized smooth nuclei at $b=4$ fm}
\label{sec:Cu}

As a warm-up example, consider the case of smooth optical Glauber initial conditions for collisions of $^{63}$Cu nuclei at an impact parameter of $b=4$ fm. In this case the initial conditions are simple enough that the main physics similarities and differences between hydro and non-interacting gas can be understood.

\begin{figure}[t]
%\centralizing
\includegraphics[width=0.45\linewidth]{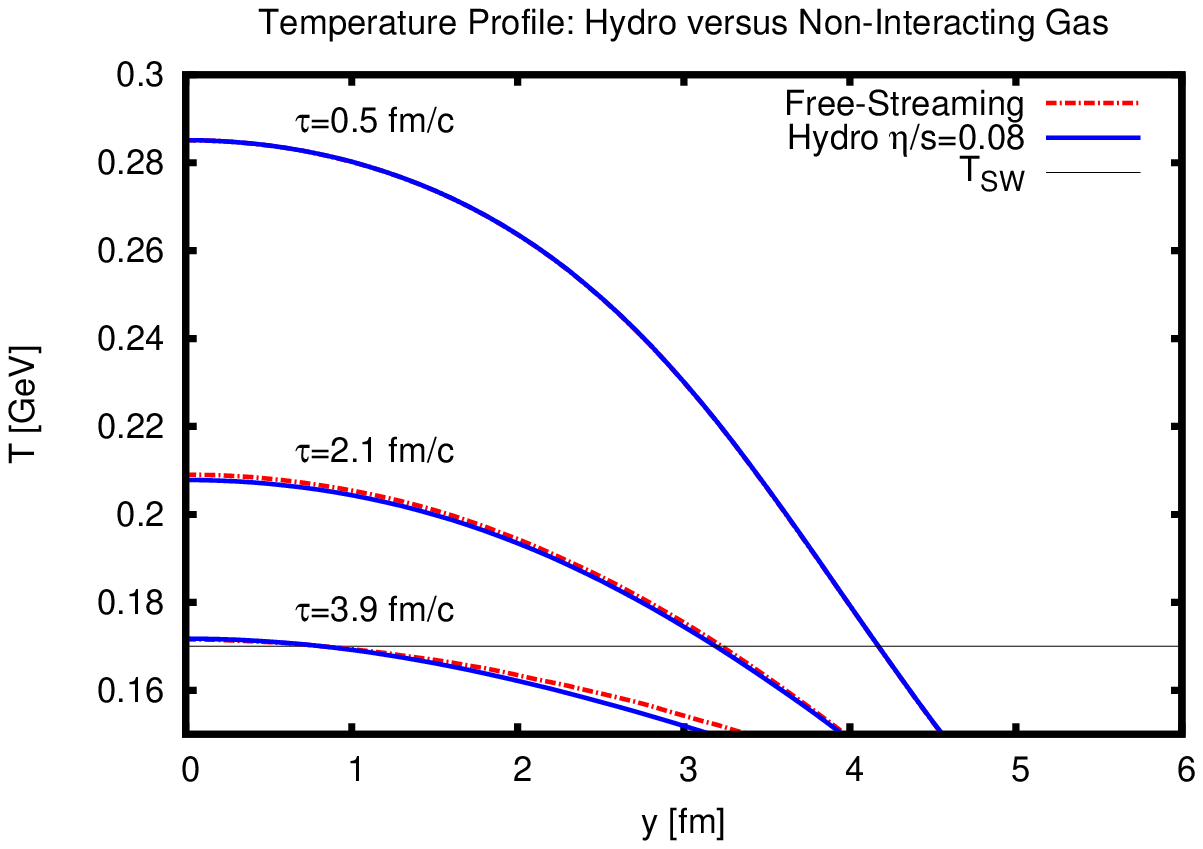}
\hfill
\includegraphics[width=0.45\linewidth]{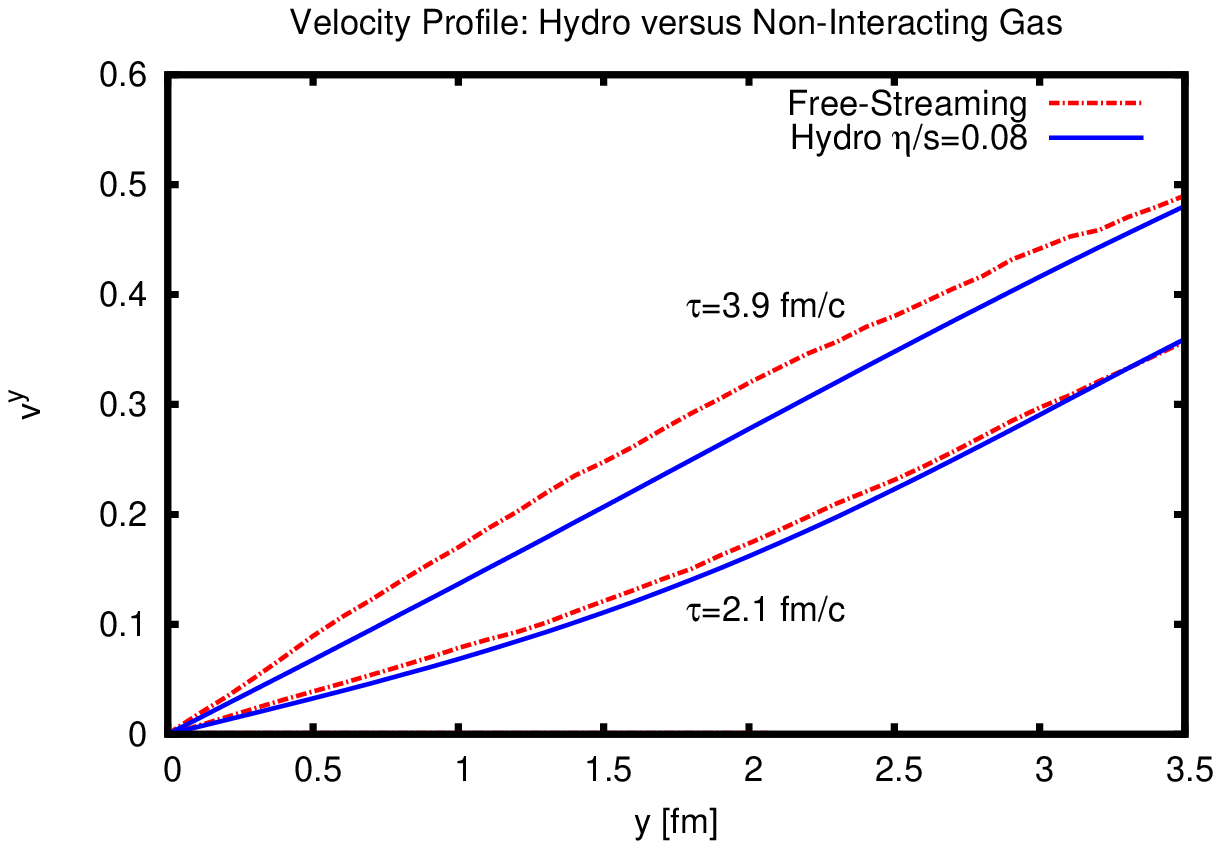}
\hfill
\includegraphics[width=0.45\linewidth]{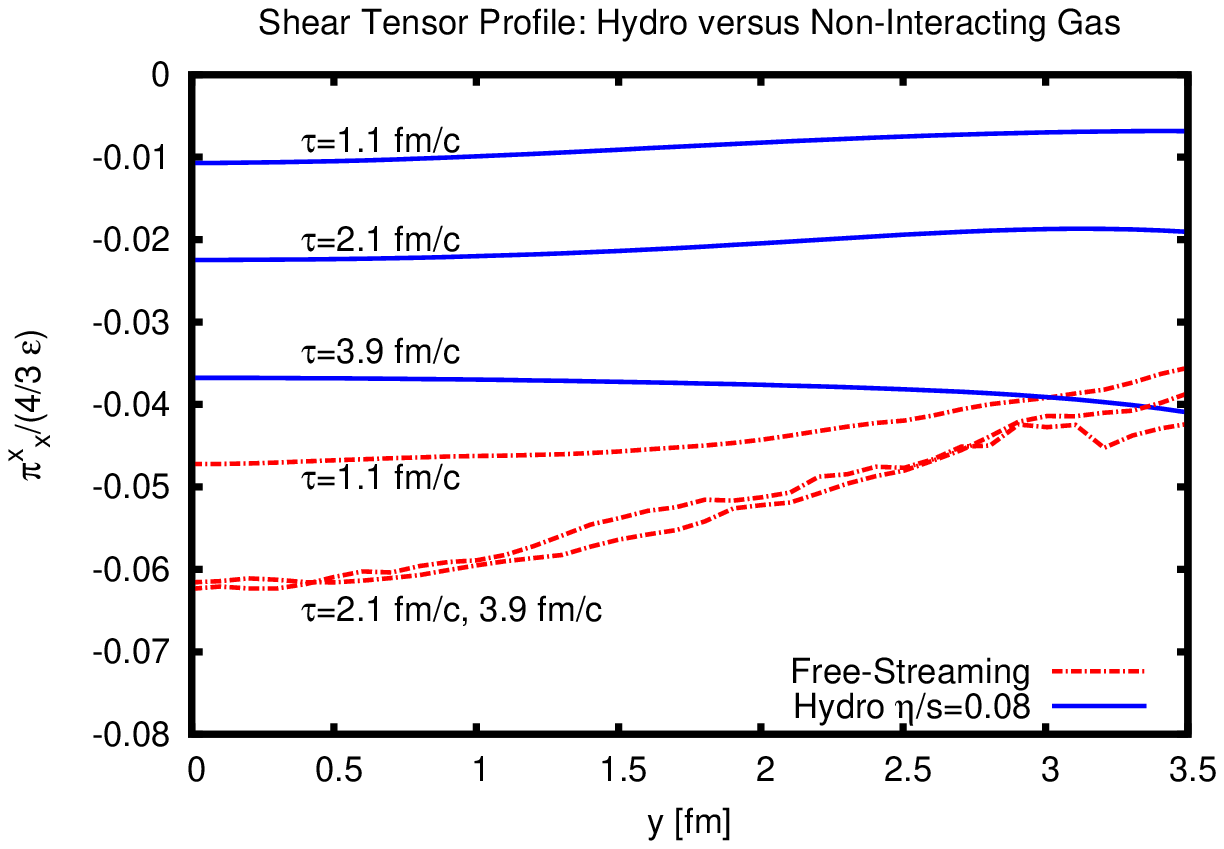}
\hfill
\includegraphics[width=0.45\linewidth]{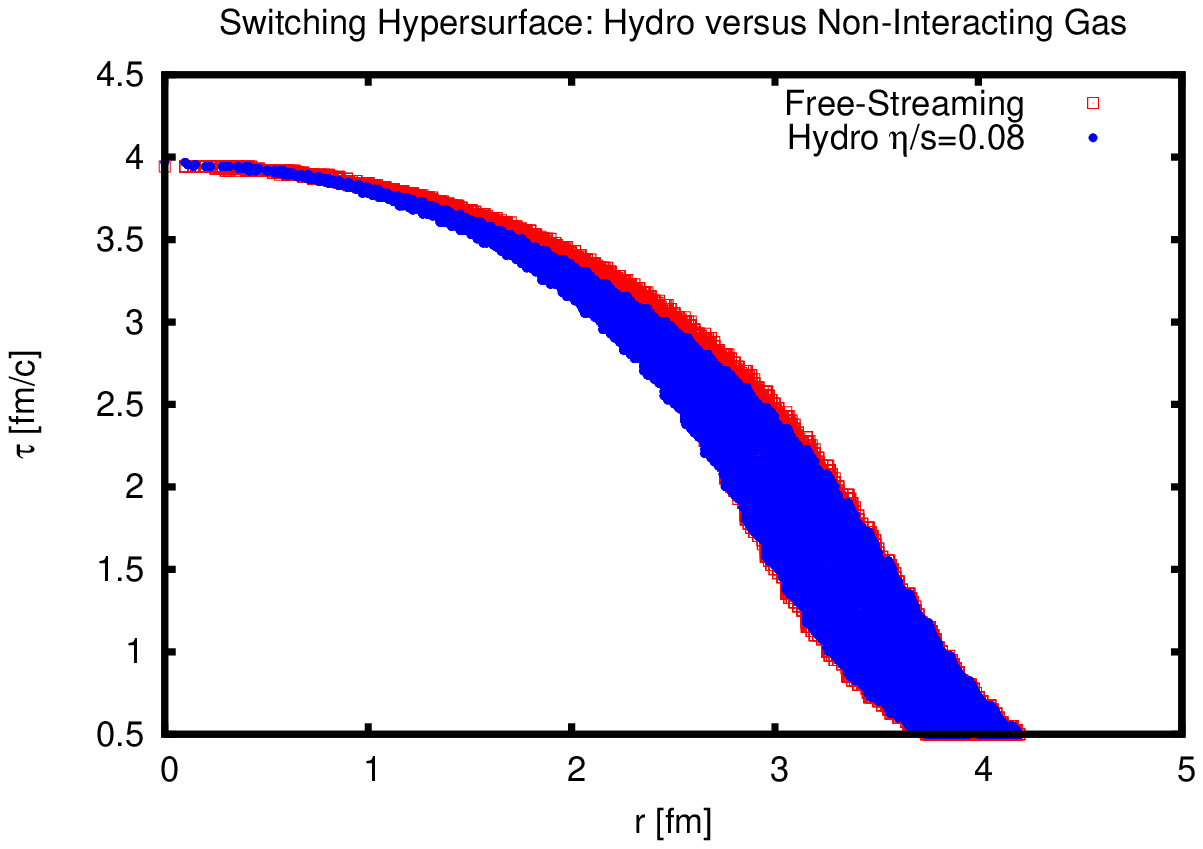}
\hfill
\includegraphics[width=0.45\linewidth]{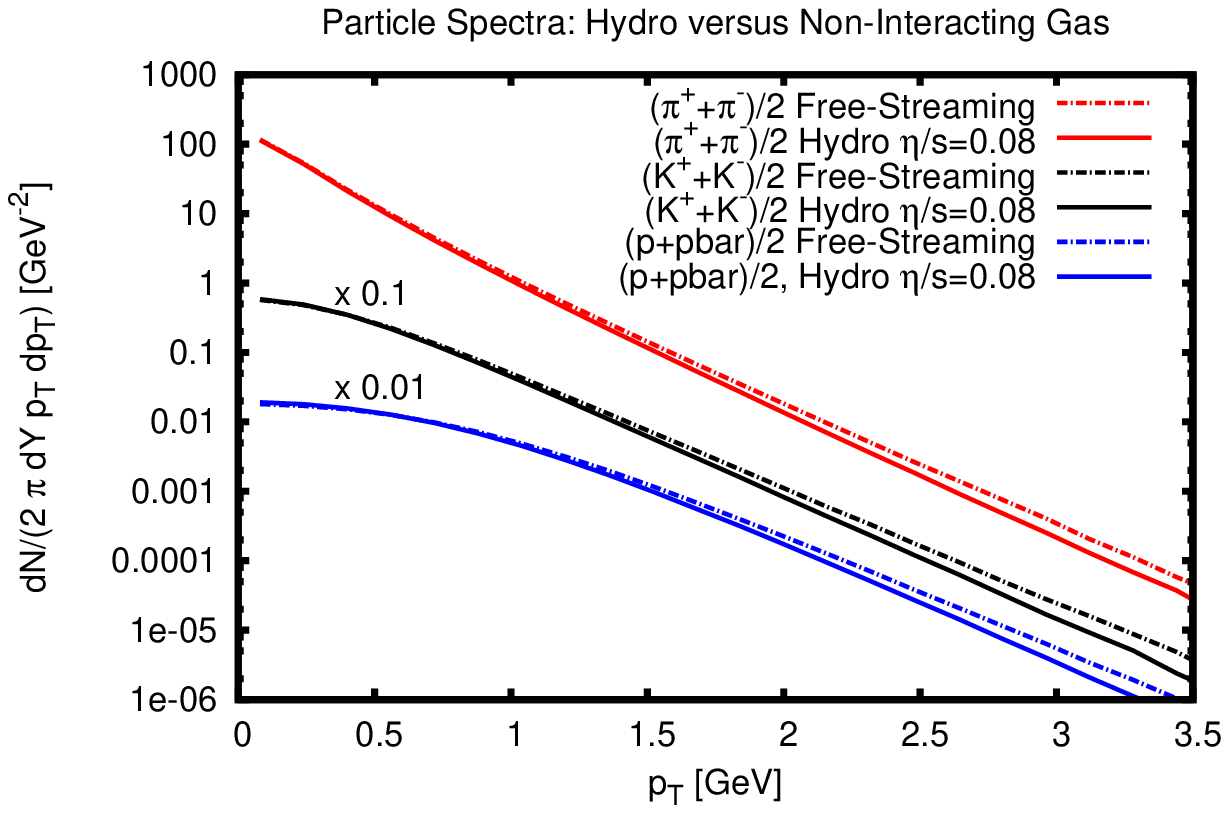}
\hfill
\includegraphics[width=0.45\linewidth]{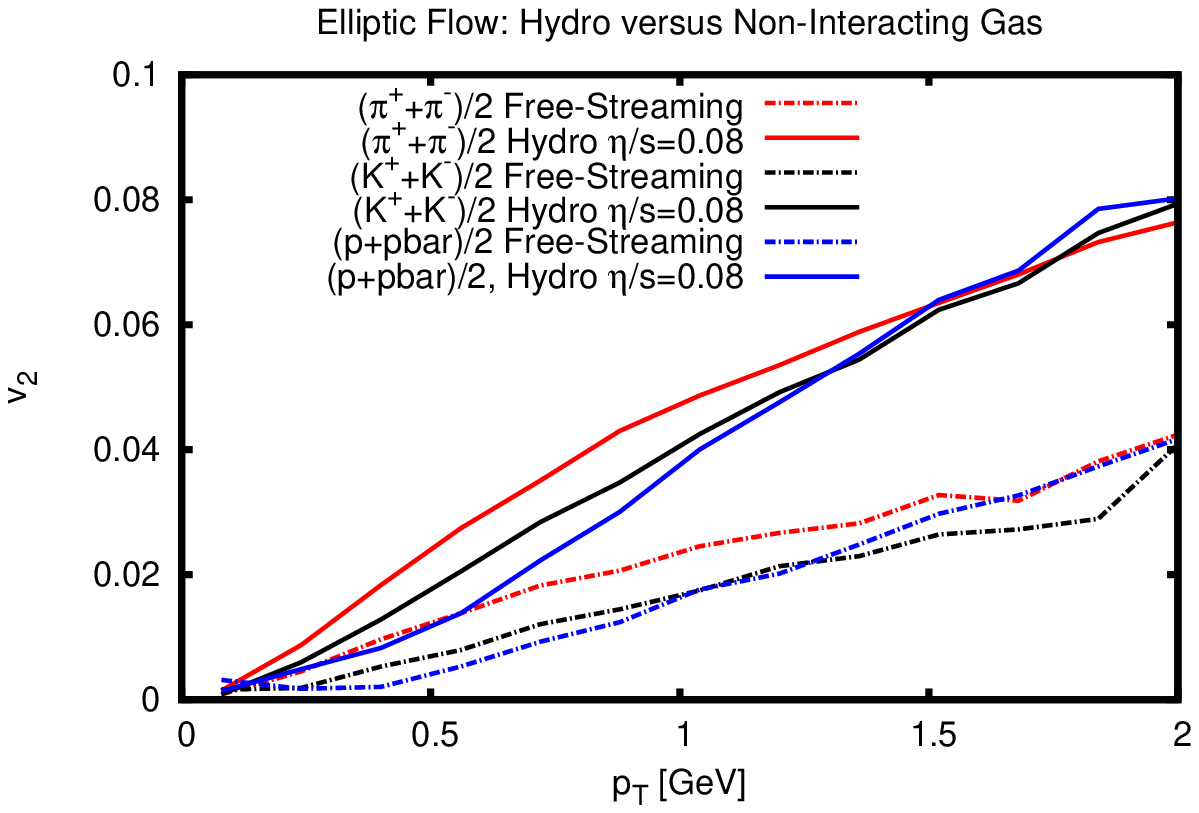}

\caption{\label{fig:two} Free-streaming evolution (no-interaction) and almost ideal hydrodynamics ($\eta/s=0.08$), followed by a hadronic cascade for smooth \CuCu collisions at $b=4$ fm. Shown are time-snapshots of the temperature profile (upper left), velocity profile (upper right), shear tensor profile (middle left), switching hypersurface (middle right), identified particle spectra (lower left) and identified particle elliptic flow (lower right). For reference, the switching temperature is indicated in the upper left plot. Note that the 'true' theoretical $v_2(p_T)$ curves would be smooth, but the finite sampling statistics for the hadron cascade code introduces some statistical error that is particularly evident for the smaller values encountered in the free-streaming case. See text for details. }
\end{figure}

In Fig.~\ref{fig:two}, time-snapshots of the temperature, velocity and shear tensor space profiles along the $y$-axis are shown. One notes that despite the very different character of the hydrodynamic and free-streaming evolution, the temperature profiles during most of the evolution are almost identical. The equal-time velocity comparison shows that flows are also similar in magnitude, but the velocities from non-interacting evolution are consistently larger than those from hydrodynamics. This is easy to explain: in almost ideal hydrodynamics, the pressure along the transverse axes $P_T$ and the longitudinal axis $P_L$ are almost identical (after all, hydrodynamics implies that the system is locally approximately isotropic). By contrast, in the free-streaming evolution in the boost-invariant approximation the longitudinal pressure falls quicker than the transverse pressure because there are no particle interactions to keep the system locally isotropic. Since the sum of the transverse and longitudinal pressure is fixed by the equation of state, this implies that the transverse pressure in free-streaming will generally be larger than the transverse pressure in almost ideal hydrodynamics. Since flow velocities in the transverse plane are being sourced by the gradient of the pressure, larger transverse pressures lead to larger flow velocities in the non-interacting case. For a particular case where the similarity of radial flow in free-streaming dynamics and hydrodynamics can be analytically demonstrated see the discussion in Appendix \ref{sec:app}.

When considering the result for the shear stress tensor in Fig.~\ref{fig:two}, the above similarities between hydrodynamic and free-streaming evolution stop. While the shear stress is generated gradually in hydrodynamics, the non-interacting free-streaming evolution leads to a sudden build-up and subsequent saturation of the shear stress. However, it is interesting to note that despite the fact that the free-streaming evolution literally corresponds to infinite viscosity, the overall magnitude of the shear stress tensor generated during the evolution is comparable to that from hydrodynamics with extremely small viscosity over entropy ratio $\eta/s\sim 0.08$.

The information about the components of the energy-momentum tensor is imprinted onto the final particle spectra, shown also in Fig.~\ref{fig:two}. It is important to recall that the final particle spectra result from either hydrodynamics or free-streaming dynamics in the hot phase of the evolution $T>T_{SW}$ followed by the same hadronic cascade evolution in the cold phase $T<T_{SW}$. Shown are results for pions, kaons and protons, and the larger transverse flow developed in the free-streaming evolution (as compared to hydrodynamics) is clearly seen as a flattening of the spectra for all particle species. From this figure, it is evident that radial flow does not indicate the presence of a hydrodynamic phase during the system evolution. This has been noticed before \cite{vanderSchee:2013pia}. Nevertheless, it is worth stressing that the presence of radial flow should not be used as an indicator for hydrodynamics, as has been often assumed (see e.g. Ref.~\cite{Kalaydzhyan:2015xba}).

Also shown in Fig.\ref{fig:two} is the identified particle elliptic flow $v_2(p_T)$ resulting from hydrodynamics or free-streaming dynamics. Note that in the simple case of smooth initial conditions, by symmetry this is the only non-trivial anisotropic flow component $v_n$. From this plot it is evident that almost ideal hydrodynamics gives rise to a considerably larger elliptic flow than free-streaming dynamics, confirming the predominant view that hydrodynamics is necessary to explain strong anisotropic flow. However, the elliptic flow found for the case of free-streaming is not consistent with zero. At first glance, this is puzzling, given that it can be analytically shown that free-streaming does not generate momentum anisotropies by itself, while diluting the spatial anisotropies (see e.g. Refs.~\cite{Kolb:2000sd,Luzum:2008cw}). However, even though the spatial anisotropies are being diluted, their potential to generate momentum anisotropies is not actually lost. Rather, what is happening is that both macroscopic velocities and dissipative parts of the stress tensor are being generate in precisely such a way that the net (non-equilibrium) momentum anisotropy is exactly zero. This is demonstrated explicitly in Appendix \ref{app:two}. In the case at hand, the full energy-momentum tensor after the free-streaming evolution is used to initialize the late-stage hadronic evolution, and it turns out that the hadronic interactions are sufficient to re-generated part of the momentum anisotropies from this $T^{ab}$ by strongly damping the dissipative parts while the flow velocities remain. This is only possible if the hadronic evolution itself has transport properties similar to a ``low'' viscosity fluid, because otherwise momentum anisotropies on the level seen in Fig. \ref{fig:two} (e.g. 50 percent of hydrodynamics with $\eta/s=0.08$) would never be (re-)generated. Recent measurements of the ratio of shear viscosity over entropy density in the hadron gas phase are consistent with this picture \cite{Romatschke:2014gna}.

The presence of a hadron gas phase (often referred to as ``corona'' in earlier work \cite{Hirano:2005wx}) is essential for generating the anisotropic flow effects seen in the identified particle plots in Fig~.\ref{fig:two}. Without hadron gas phase, there would be radial flow (see the analytic result presented in Appendix \ref{sec:app}), but no elliptic flow. However, in actual systems created in relativistic ion collisions there always is a hadronic gas phase, so it is crucial that this component be included in the system description, and that its transport properties are better quantified (see e.g. Ref.~\cite{Romatschke:2014gna} for work along these lines).

\section{Results}
\label{sec:res}

\begin{figure}[t]
%\centralizing
\includegraphics[width=0.45\linewidth]{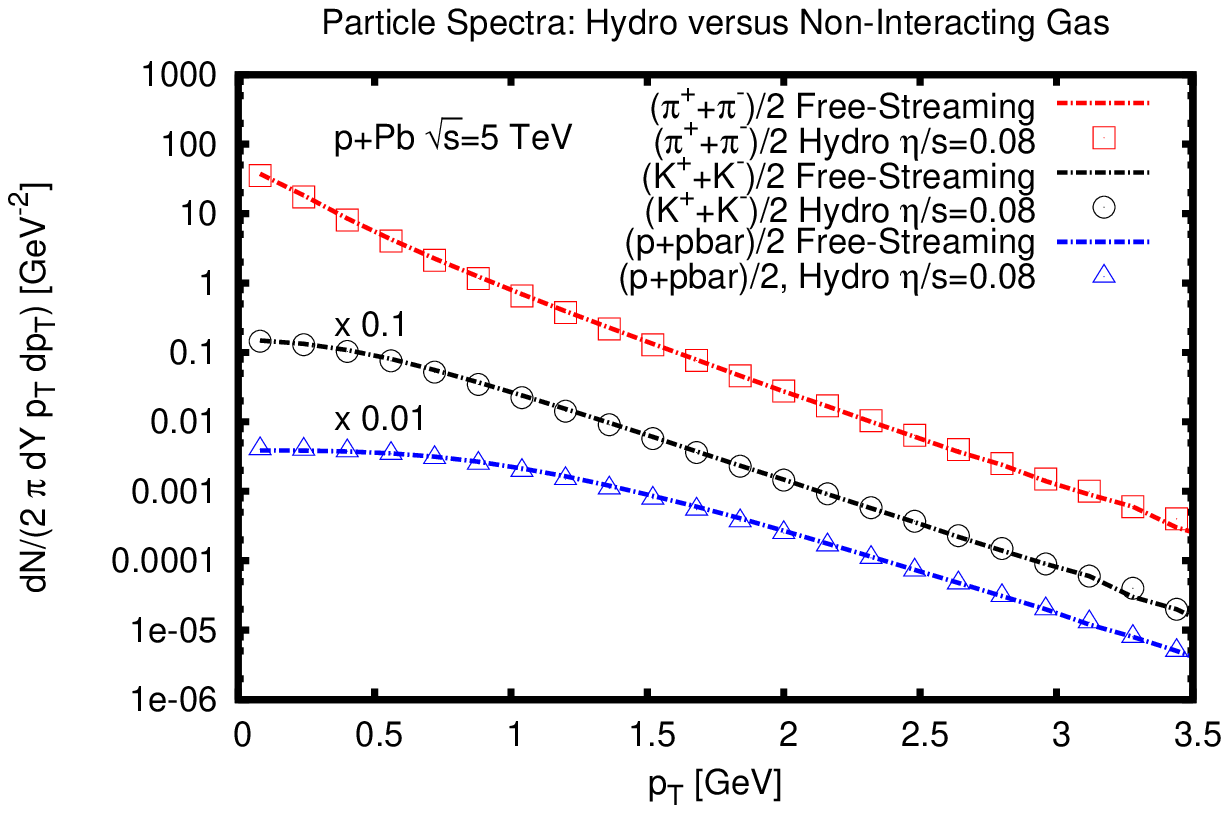}
\hfill
\includegraphics[width=0.45\linewidth]{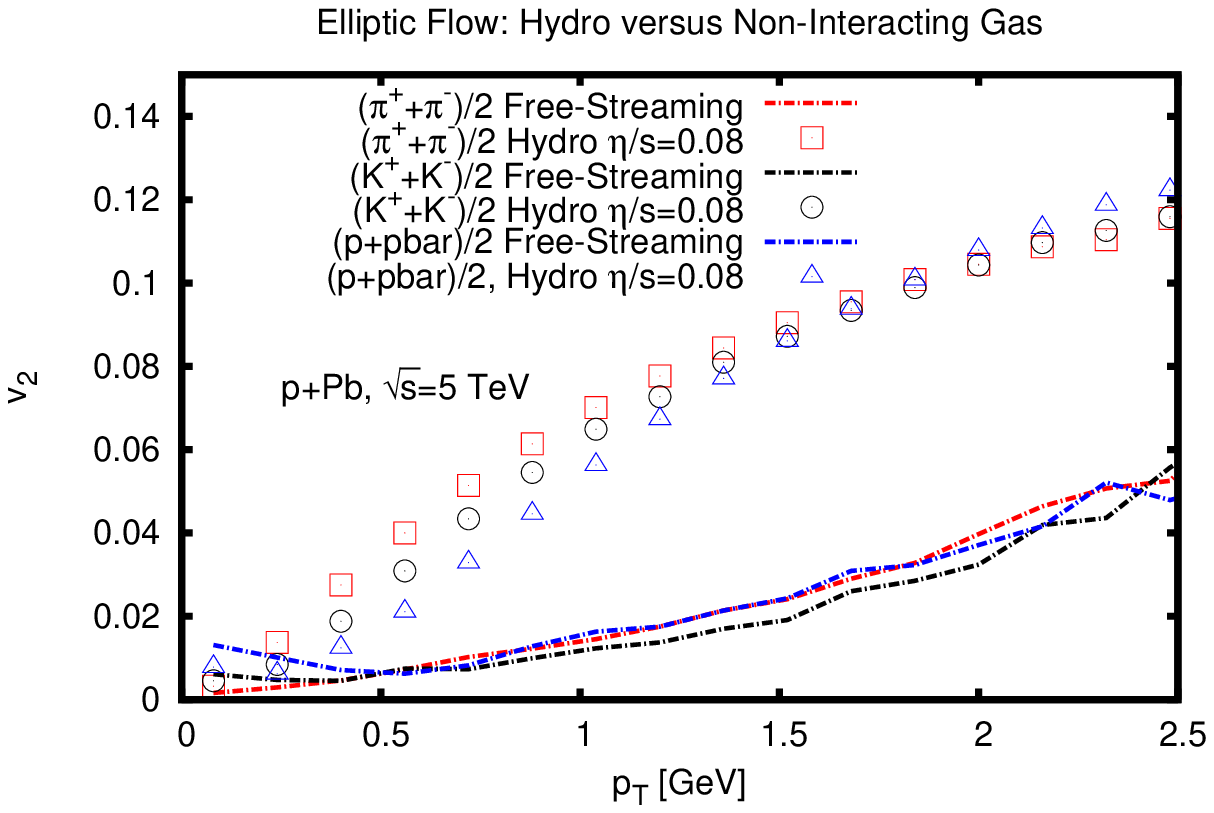}
\hfill
\includegraphics[width=0.45\linewidth]{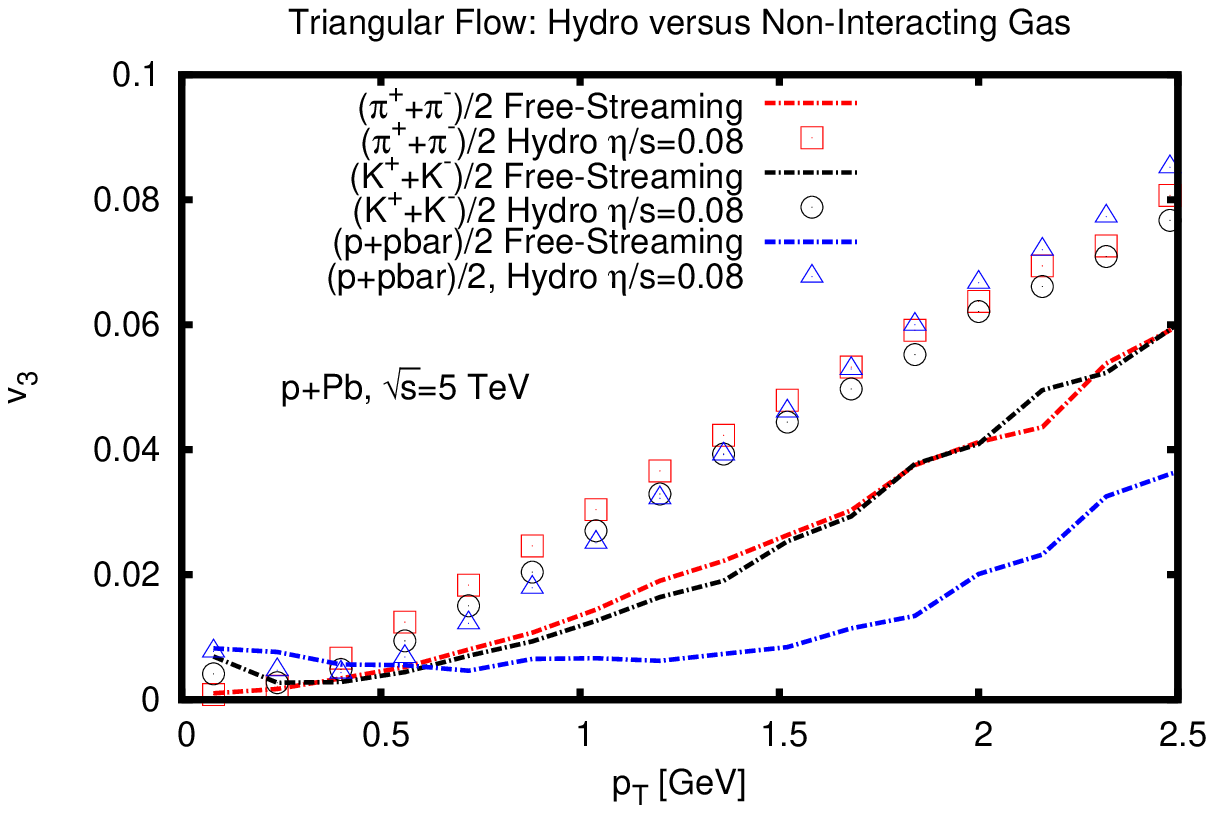}
\hfill
\includegraphics[width=0.45\linewidth]{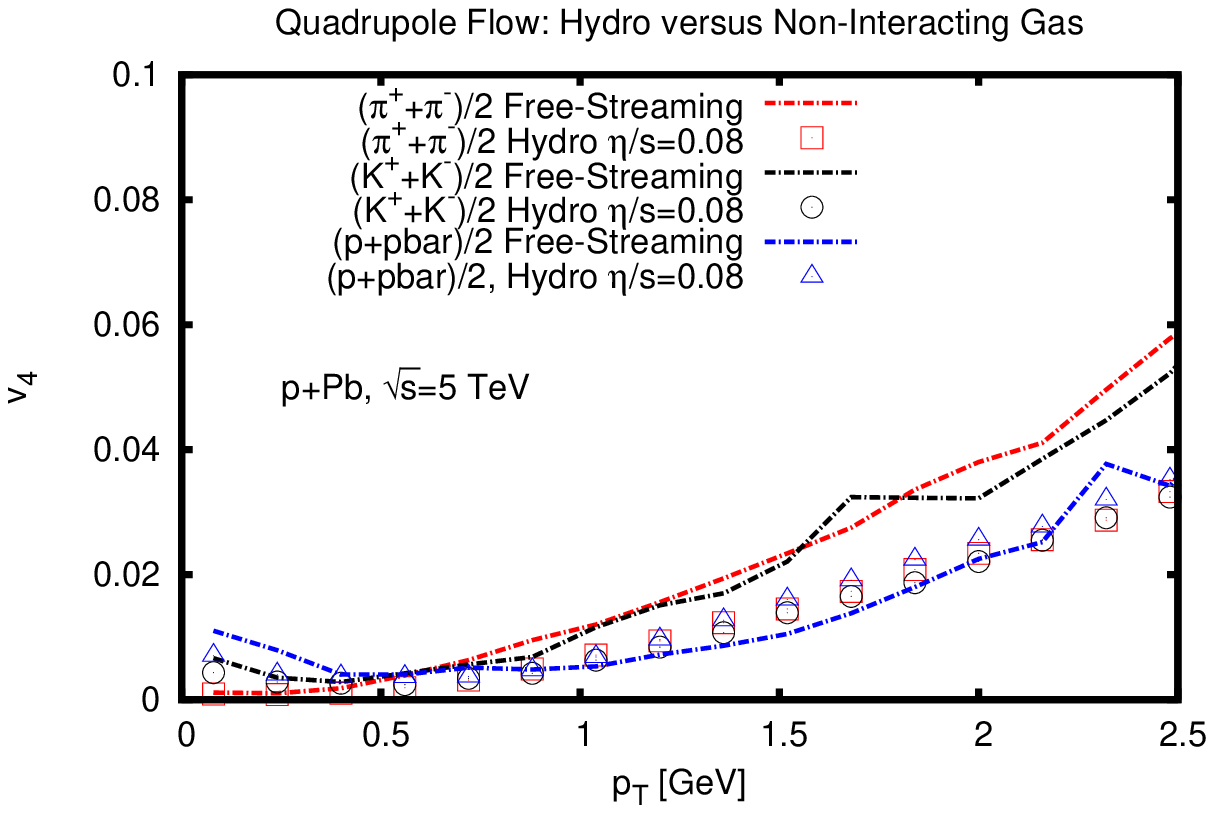}
\caption{\label{fig:three} Simulations of granular \pPb collisions at $\sqrt{s}=5$ TeV. Shown are final particle spectra and anisotropic flow coefficients $v_n(p_T)$ for identified particles for free-streaming evolution (no-interaction) and almost ideal hydrodynamics ($\eta/s=0.08$), followed by a hadronic cascade.  See text for details. }
\end{figure}

\subsection{Central event-by-event collisions of granular nuclei}

A more realistic application of the techniques highlighted in the previous section is the relativistic collision of light-on-heavy-ions, such as \hAu and \dAu at $\sqrt{s}=200$ GeV and \pPb at $\sqrt{s}=5.02$ TeV. For each of these collision systems, one hundred initial events are generated from a probability distribution of nucleons inside the colliding nuclei (see section \ref{sec:IS} for a detailed discussion). For each of these events, the subsequent dynamics is simulated using either a hydrodynamic evolution or a free-streaming evolution, followed by the same hadron cascade for the low temperature phase. Unlike the simplified case discussed in the section \ref{sec:Cu}, the granular nature of each individual event gives rise to all anisotropic flow harmonics $v_n$ with $n\geq1$, not just the elliptic flow $v_2$.

The results for \pPb collisions at $\sqrt{s}=5.02$ TeV are shown in Fig.~\ref{fig:three}. Considering the identified particle spectra, one finds that the additional radial flow generated in the free-streaming dynamics compared to hydrodynamics is almost negligible, and the resulting spectra are essentially indistinguishable. One reason for this may be the comparatively shorter evolution time spent in the hot phase $T>T_{SW}$ for \pPb collisions compared to the case of smooth nucleus-nucleus collisions considered in Sec.\ref{sec:Cu}. 

The comparison between free-streaming dynamics and hydrodynamics for the elliptic flow coefficient $v_2$ are consistent with the findings for smooth nucleus-nucleus collisions considered above: the coupled free-streaming and hadron gas dynamics gives rise to a non-negligible amount of $v_2$, but it is considerably less than the $v_2$ generated in hydrodynamics.

Considering the higher flow harmonics $v_3,v_4$, the comparison between free-streaming and hydrodynamics reveals that it becomes more difficult to distinguish between the two scenarios in terms of flow magnitude. For instance, the $v_3$ found for free-streaming plus hadron cascade dynamics is very similar in magnitude to that for hydrodynamics plus hadron cascade. Maybe even more interesting, the $v_4$ amplitude for the free-streaming plus cascade simulation in \pPb collisions turns out to be \emph{larger} than the corresponding result from hydrodynamics with $\eta/s=0.08$ (see Fig.~\ref{fig:three}). 

Overall one finds that free-streaming dynamics followed by hadron cascade dynamics generates approximately the same magnitude of anisotropic flow for $v_2,v_3$ and $v_4$, e.g. independent from the order of the harmonic. This is clearly very different from hydrodynamics, where successively higher orders are more strongly suppressed.

\begin{figure}[t]
%\centralizing
\includegraphics[width=0.45\linewidth]{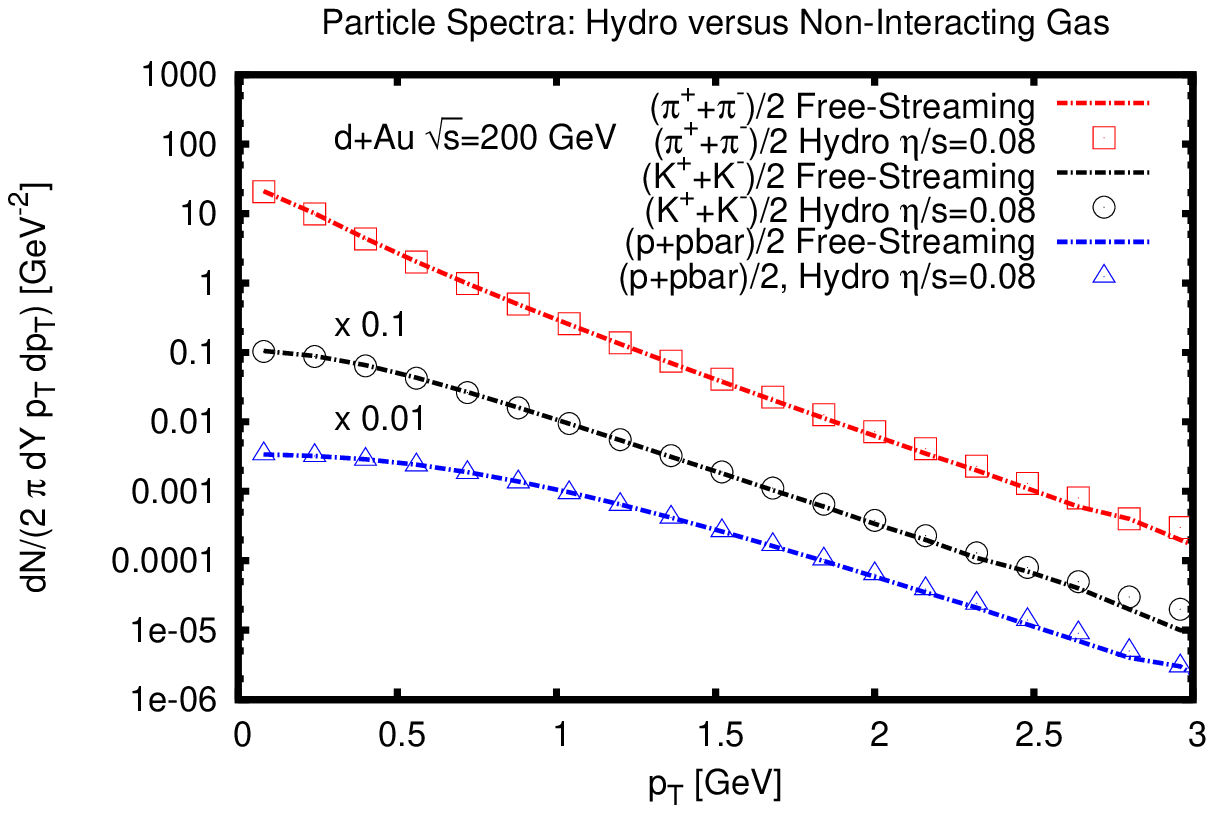}
\hfill
\includegraphics[width=0.45\linewidth]{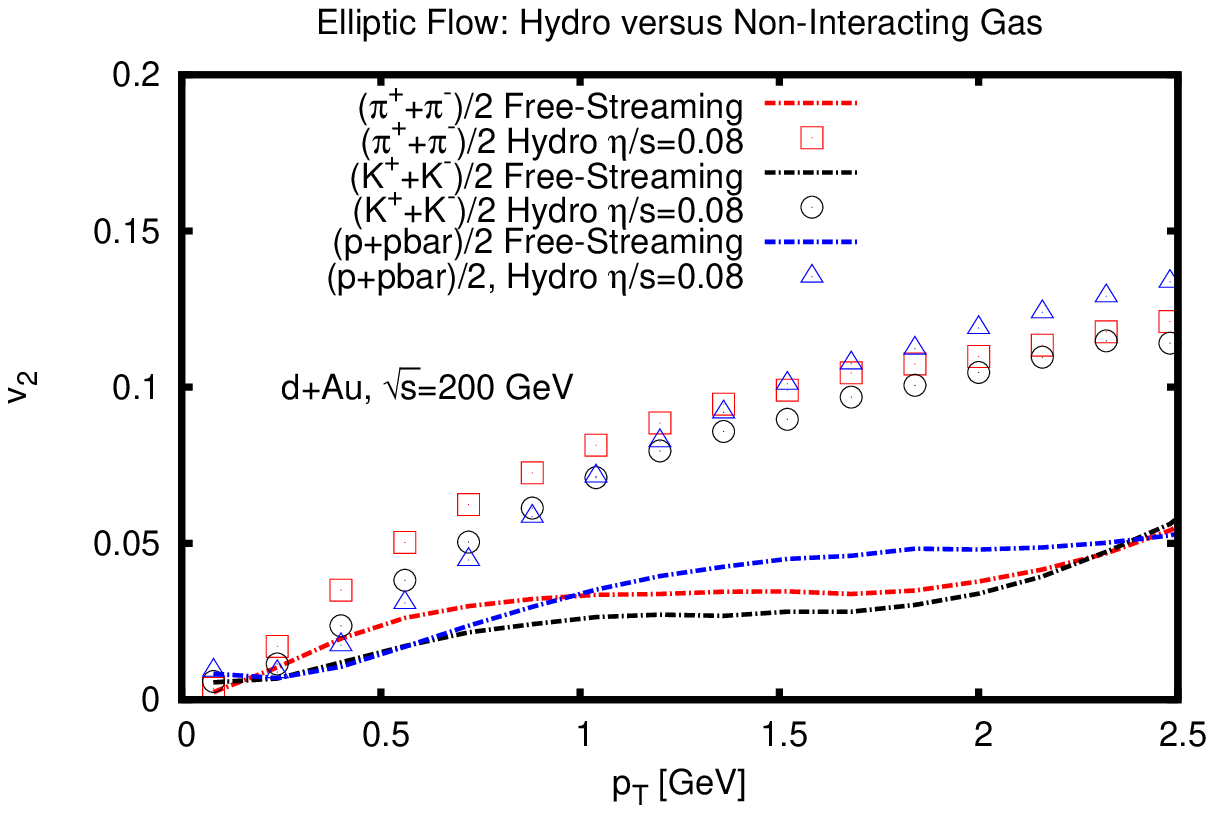}
\hfill
\includegraphics[width=0.45\linewidth]{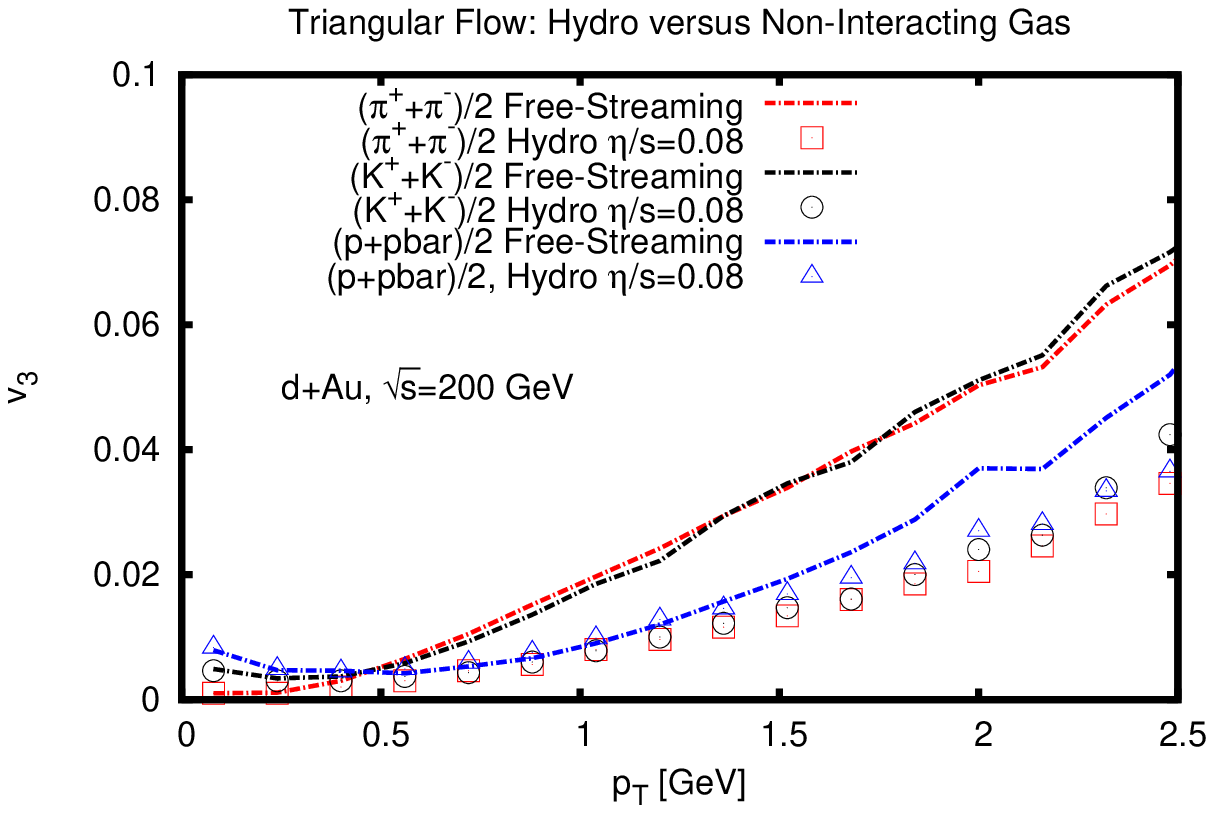}
\hfill
\includegraphics[width=0.45\linewidth]{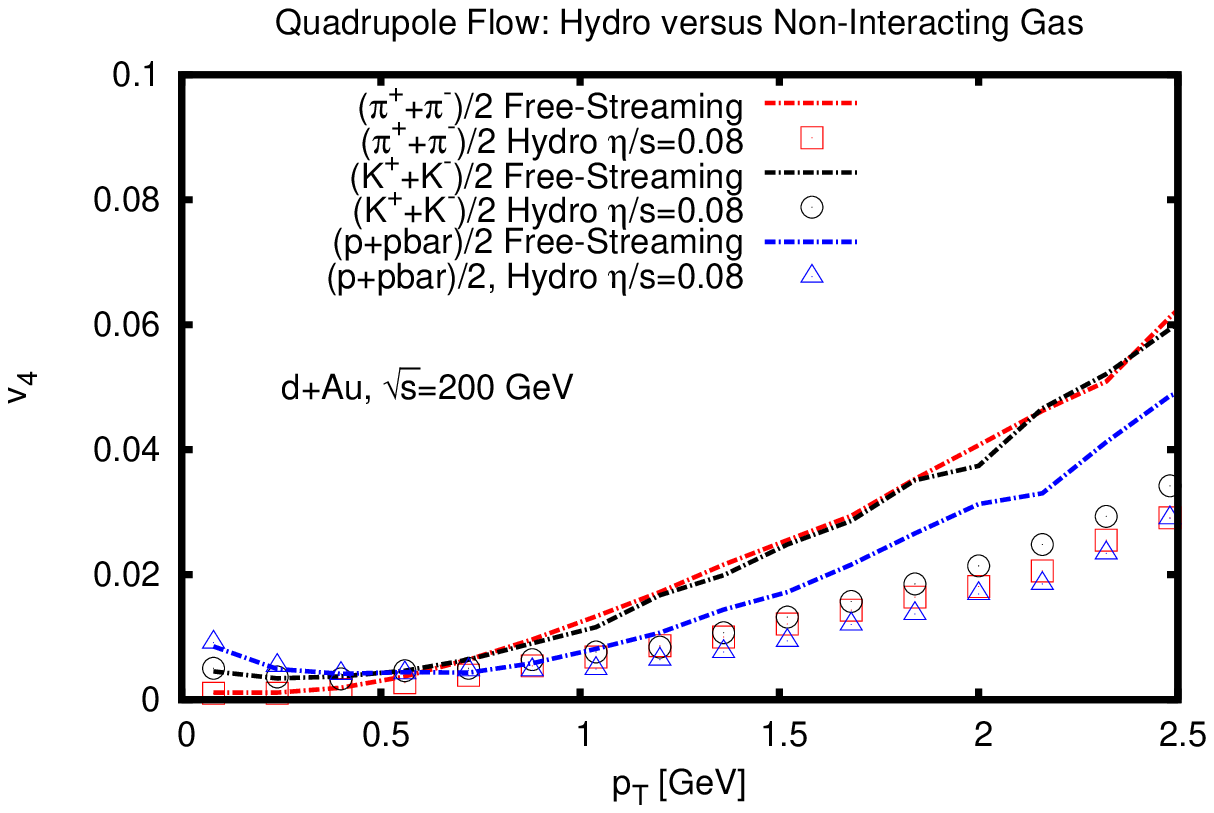}
\caption{\label{fig:four} Simulations of granular \dAu collisions at $\sqrt{s}=200$ GeV. Shown are final particle spectra and anisotropic flow coefficients $v_n(p_T)$ for identified particles for free-streaming evolution (no-interaction) and almost ideal hydrodynamics ($\eta/s=0.08$), followed by a hadronic cascade.  See text for details. }
\end{figure}

\begin{figure}[t]
%\centralizing
\includegraphics[width=0.45\linewidth]{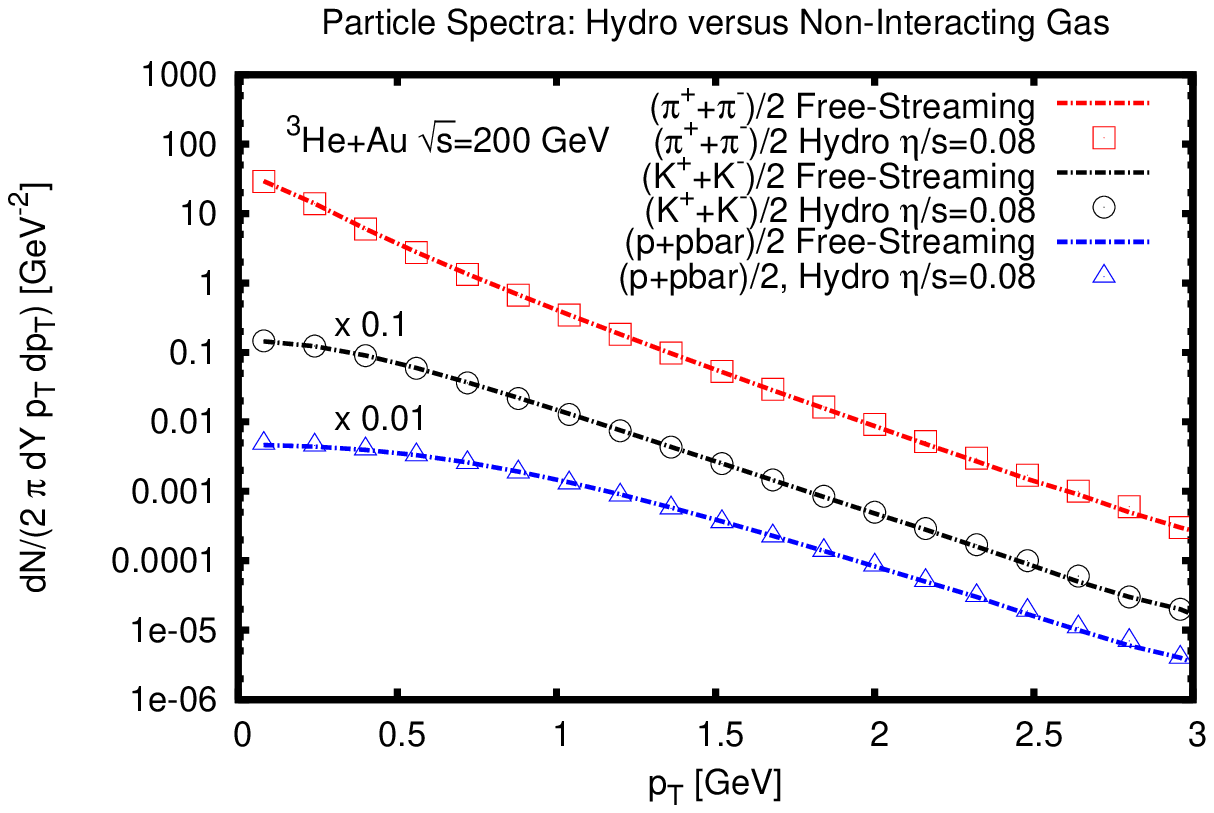}
\hfill
\includegraphics[width=0.45\linewidth]{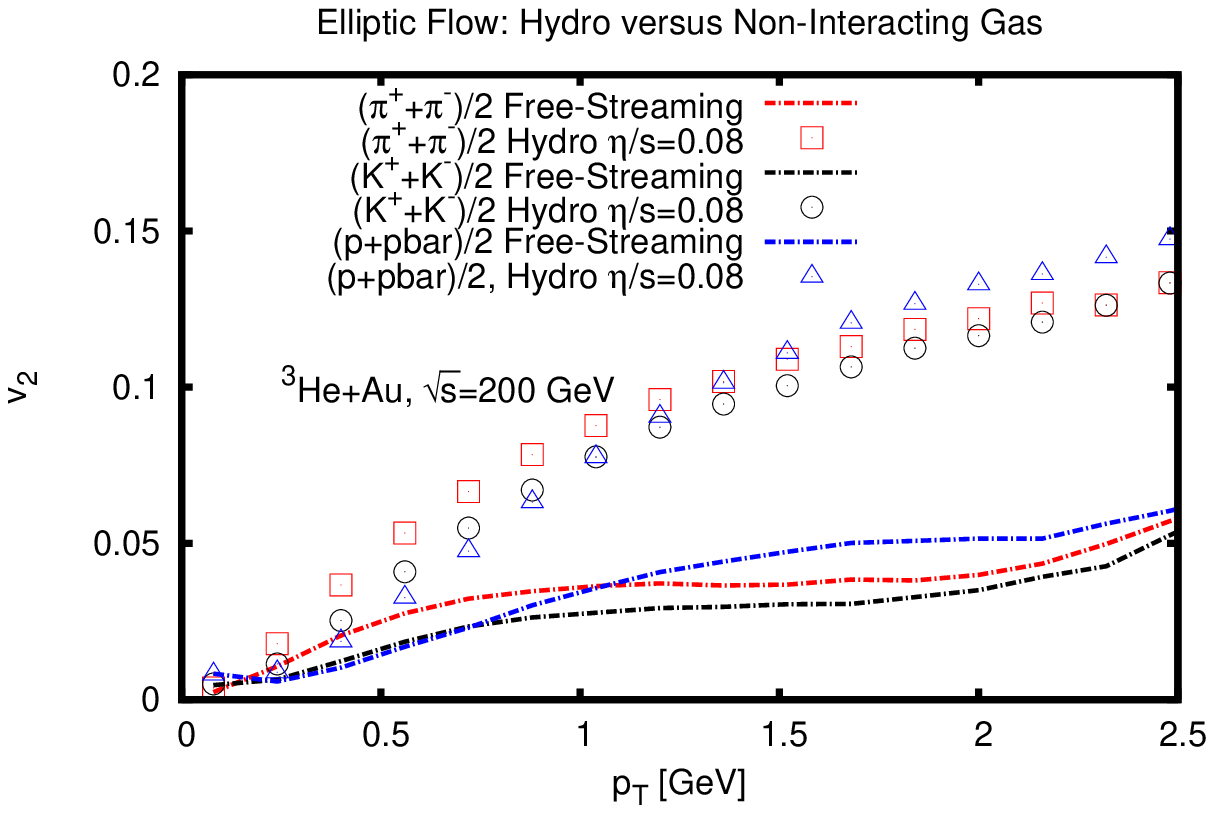}
\hfill
\includegraphics[width=0.45\linewidth]{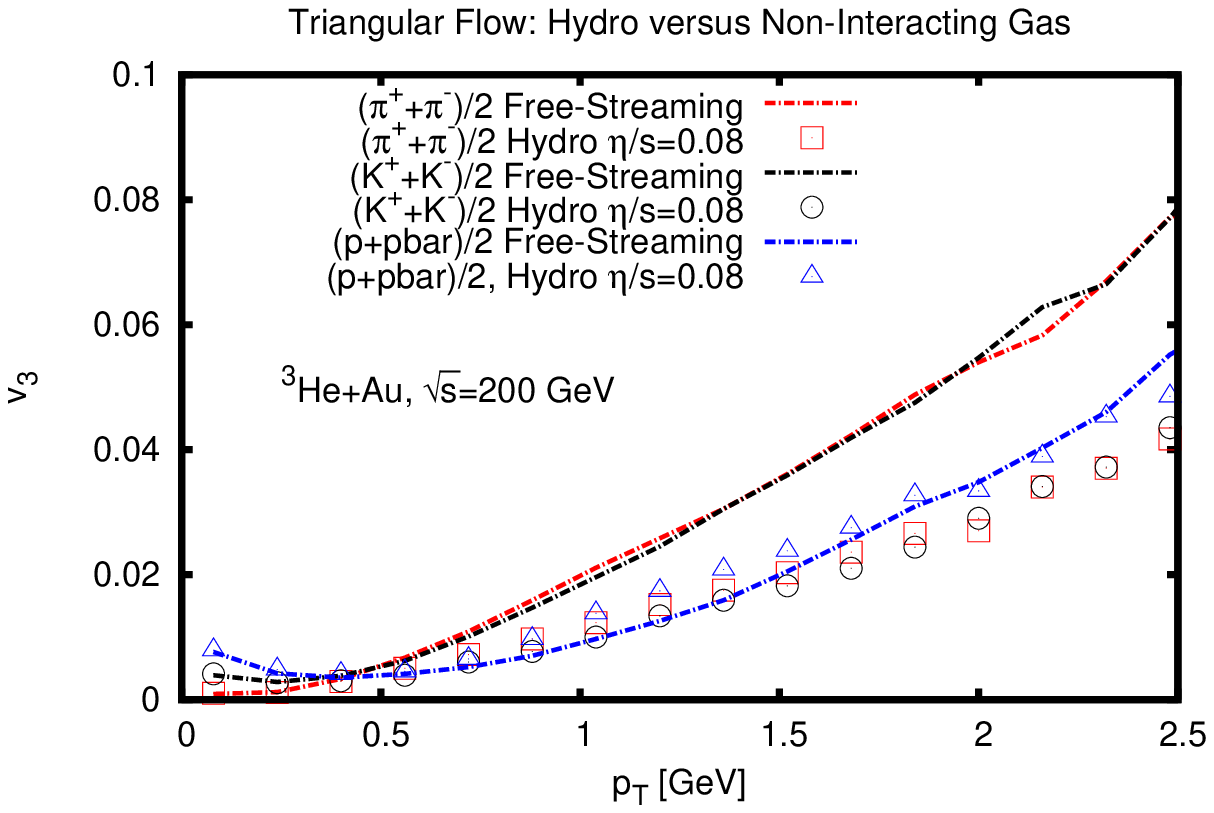}
\hfill
\includegraphics[width=0.45\linewidth]{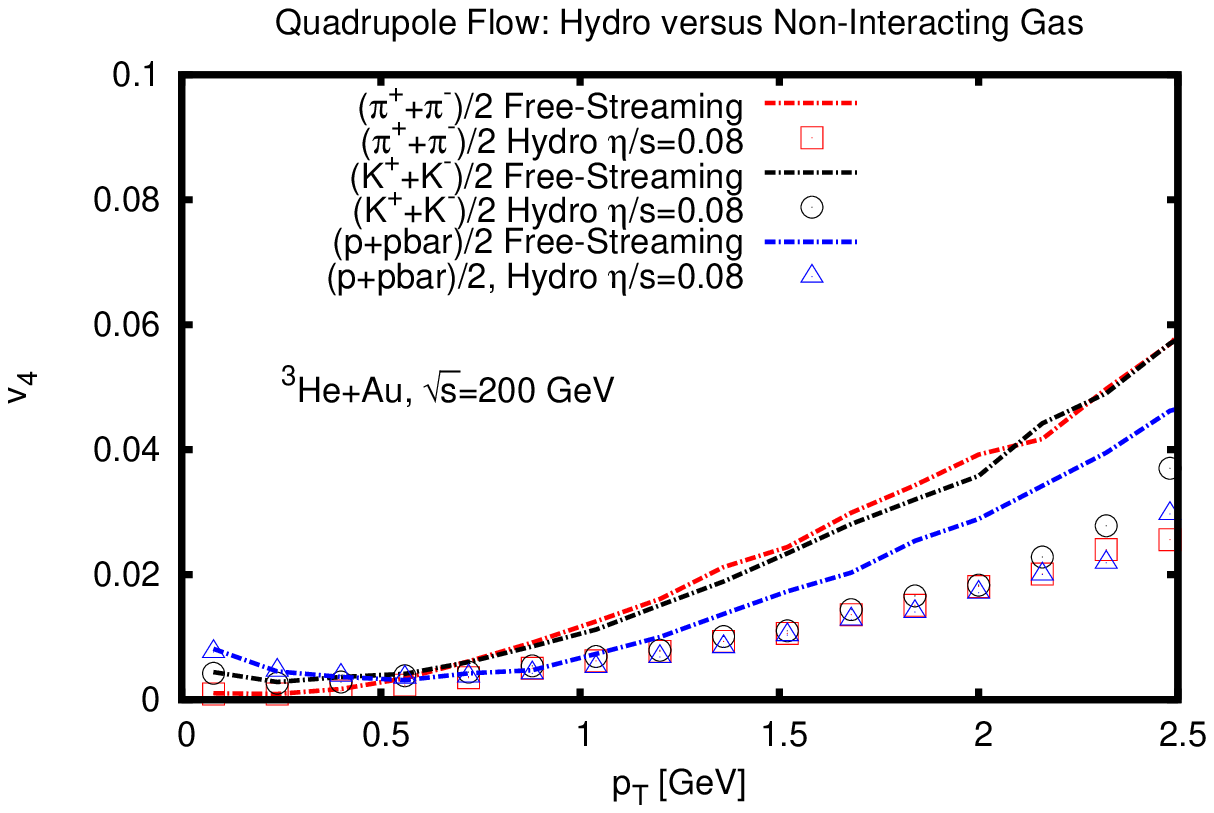}
\caption{\label{fig:five} Simulations of granular \hAu collisions at $\sqrt{s}=200$ GeV. Shown are final particle spectra and anisotropic flow coefficients $v_n(p_T)$ for identified particles for free-streaming evolution (no-interaction) and almost ideal hydrodynamics ($\eta/s=0.08$), followed by a hadronic cascade.  See text for details. }
\end{figure}

The results obtained for the \pPb collisions at $\sqrt{s}=5.02$ TeV  should be compared to the results for \dAu and \hAu collisions at $\sqrt{s}=200$ GeV energies shown in Figs.~\ref{fig:four},\ref{fig:five}. Overall, the same trends that were identified in \pPb collisions repeat for these lower-energy collision systems. However, in \dAu and \hAu collisions at $\sqrt{s}=200$ GeV one finds that free-streaming plus hadron cascade dynamics generates larger $v_3,v_4$ than hydrodynamics with $\eta/s=0.08$. Only final $v_2$ is larger in hydrodynamics than in free-streaming.

Finally, it is curious to note that the proton $v_3$ is much smaller than pion and kaon $v_3$ in free-streaming plus cascade dynamics in \pPb, \dAu and \hAu collisions compared to the case of hydrodynamics plus cascade. This could suggest a potential experimental handle on separating $v_3$ generated by hydrodynamics from $v_3$ generated by non-hydrodynamic processes such as free-streaming. It should be cautioned that simulated proton results could be unreliable because of the missing baryon annihilation process in the hadron cascade, see Sec.\ref{sec:s3}. Nevertheless, the fact that a large proton $v_3$ suppression is seen in free-streaming plus cascade but not hydrodynamics plus cascade using the same cascade code and for almost identical final identical final particle spectra seems to suggest that this effect could be robust.

\subsection{Mid-central event-by-event nucleus-nucleus collisions}

\begin{figure}[t]
%\centralizing
\includegraphics[width=0.45\linewidth]{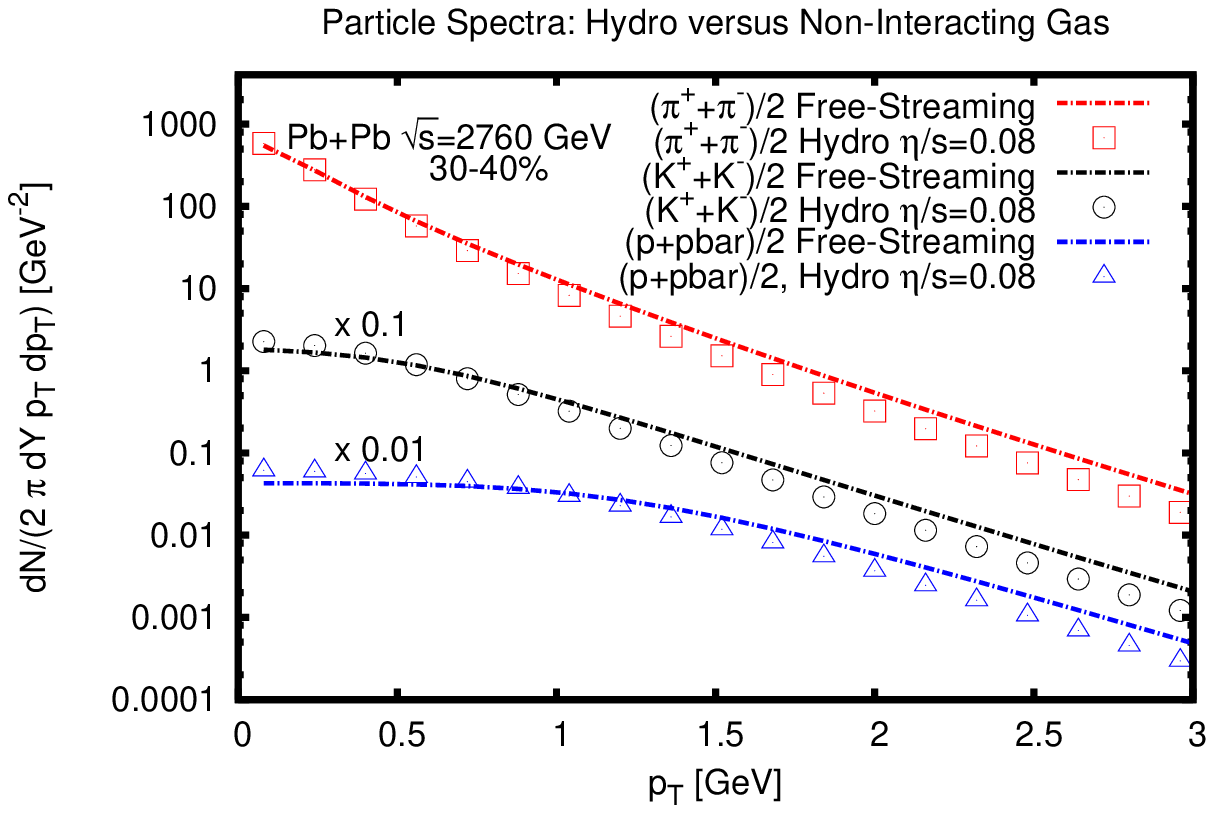}
\hfill
\includegraphics[width=0.45\linewidth]{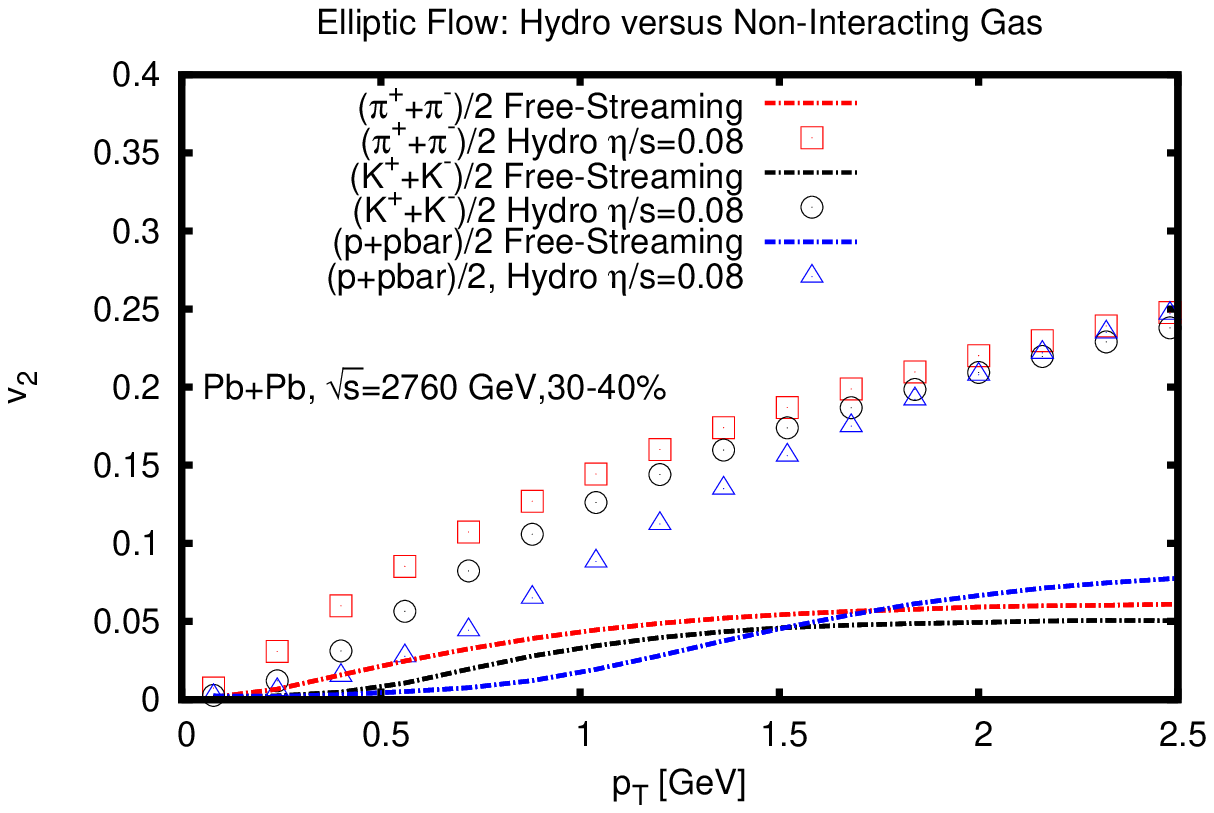}
\hfill
\includegraphics[width=0.45\linewidth]{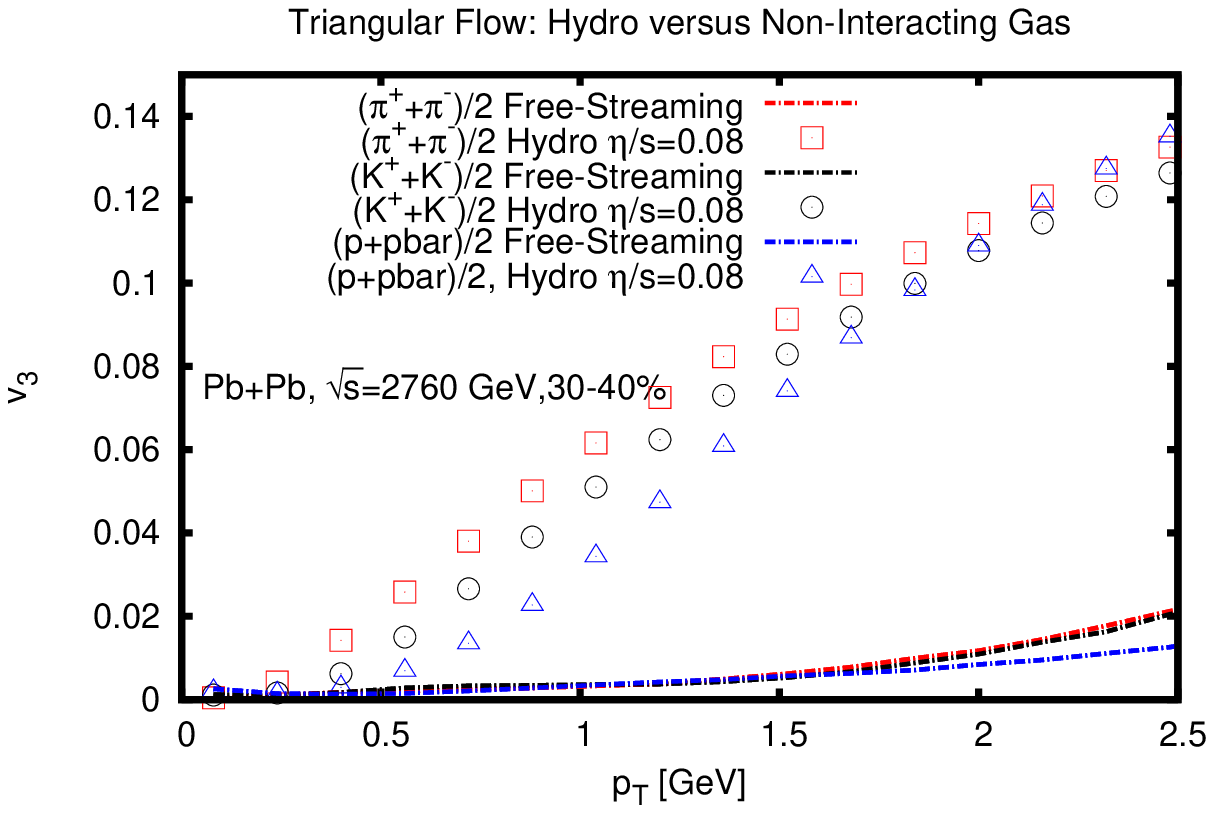}
\hfill
\includegraphics[width=0.45\linewidth]{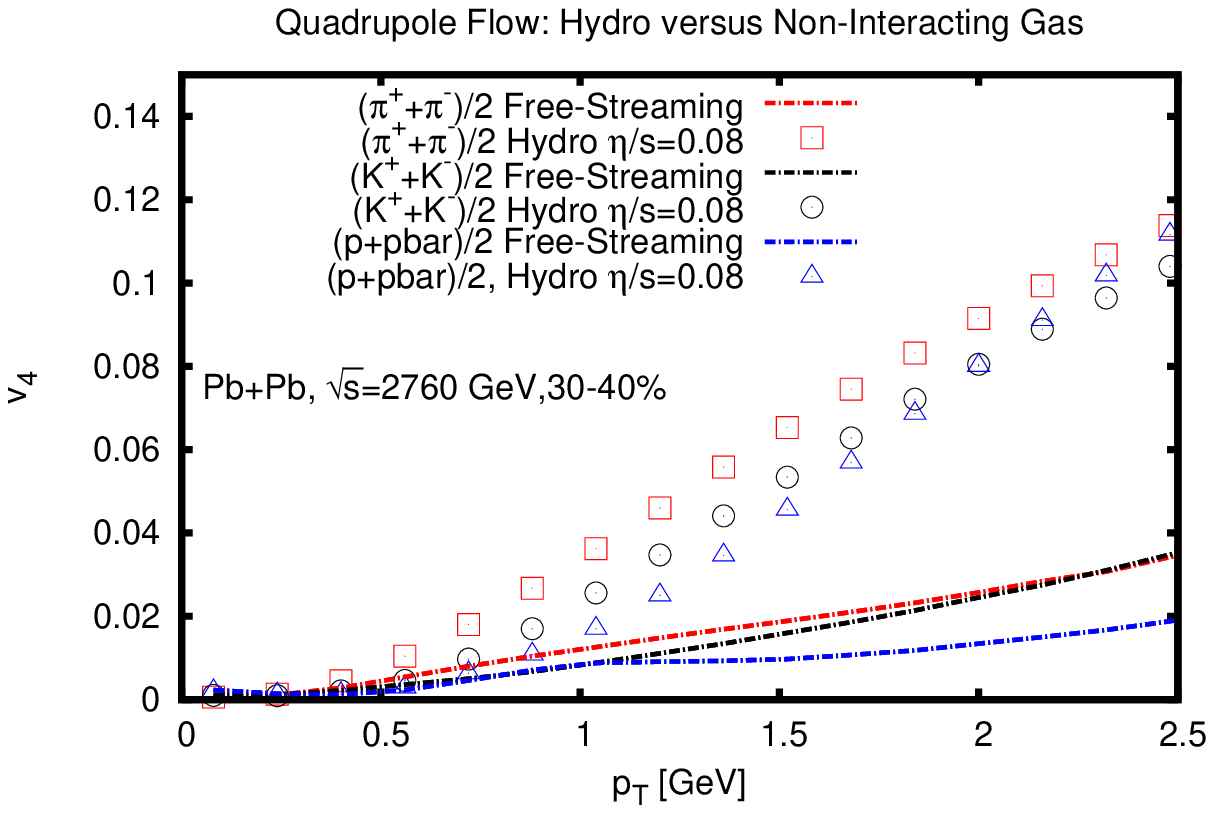}
\caption{\label{fig:PbPb} Simulations of granular \PbPb collisions at $\sqrt{s}=2.76$ TeV. Shown are final particle spectra and anisotropic flow coefficients $v_n(p_T)$ for identified particles for free-streaming evolution (no-interaction) and almost ideal hydrodynamics ($\eta/s=0.08$), followed by a hadronic cascade.  See text for details. }
\end{figure}

In view of the above findings for relativistic light-on-heavy-ion collisions, a comparison to nucleus-nucleus collisions is in order to see if similar effects are found in these larger systems. Specifically, \PbPb collisions at $\sqrt{s}=2.76$ TeV and 30\%-40\% centrality are studied. Results are shown in Fig.~\ref{fig:PbPb}. The particle spectra are broadly consistent with the warm-up case studied in section \ref{sec:Cu}, indicating a slightly larger radial flow in the free-streaming dynamics than in hydrodynamics. The elliptic flow coefficient generated in free-streaming dynamics of \PbPb collisions is found to be much smaller than the elliptic flow generated in hydrodynamics at $\frac{\eta}{s}=0.08$. For higher flow harmonics ($v_3,v_4$) the ratio between hydrodynamics and free-streaming is even larger than for $v_2$. This strongly indicates that for larger systems, the longer time spent in the hot ($T>T_{SW}$) phase prevents the free-streaming dynamics to preserve the space-anisotropies until the hadron cascade takes over, effectively leading to a strong decrease of momentum anisotropies with respect to almost ideal hydrodynamics. In essence, this confirms the established paradigm that the magnitude of anisotropic flow measured in nucleus-nucleus collisions requires a hydrodynamic phase be present during the system evolution.

One could be worried that this conclusion could be avoided by shortening the time spent in the free-streaming phase through increasing the switching temperature $T_{SW}$. However, note that for this study $T_{SW}=170$ MeV was chosen. Increasing $T_{SW}$ even further seems to not be justifiable since a hadron gas description is disfavored from lattice QCD calculations of the equation of state for $T>170$ MeV (cf. Ref.~\cite{Borsanyi:2013bia}). 

Also, because the anisotropic flow signals in hydrodynamics dwarf the free-streaming results for mid-central \PbPb collisions at $\sqrt{s}=2.76$ TeV in Fig.~\ref{fig:PbPb}, one expects qualitatively similar results for mid-central \AuAu collisions at $\sqrt{s}=200$ GeV. However, either in very low collision energy or very peripheral nucleus-nucleus collisions the time the system spends in the hot QCD phase is presumably comparable to that in central light-on-heavy-ion collisions, so that for these systems one can expect results along the lines of Figs.\ref{fig:three},\ref{fig:four},\ref{fig:five}.

\subsection{Pion Femtoscopy}

Besides flow signals, other experimentally accessible signals such as femtoscopic measurements are often used to infer the presence of a hydrodynamic phase in the evolution. 

\begin{figure}[t]
%\centralizing
\includegraphics[width=0.45\linewidth]{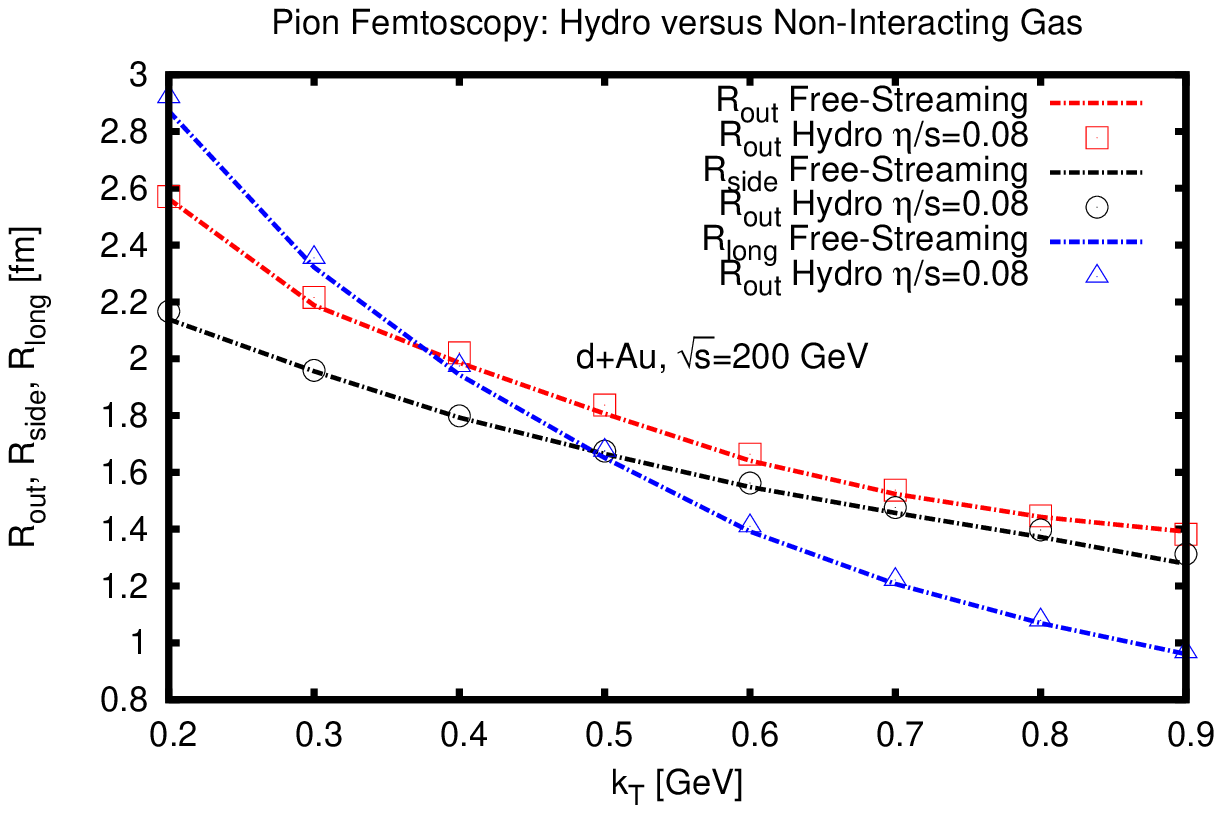}
\hfill
\includegraphics[width=0.45\linewidth]{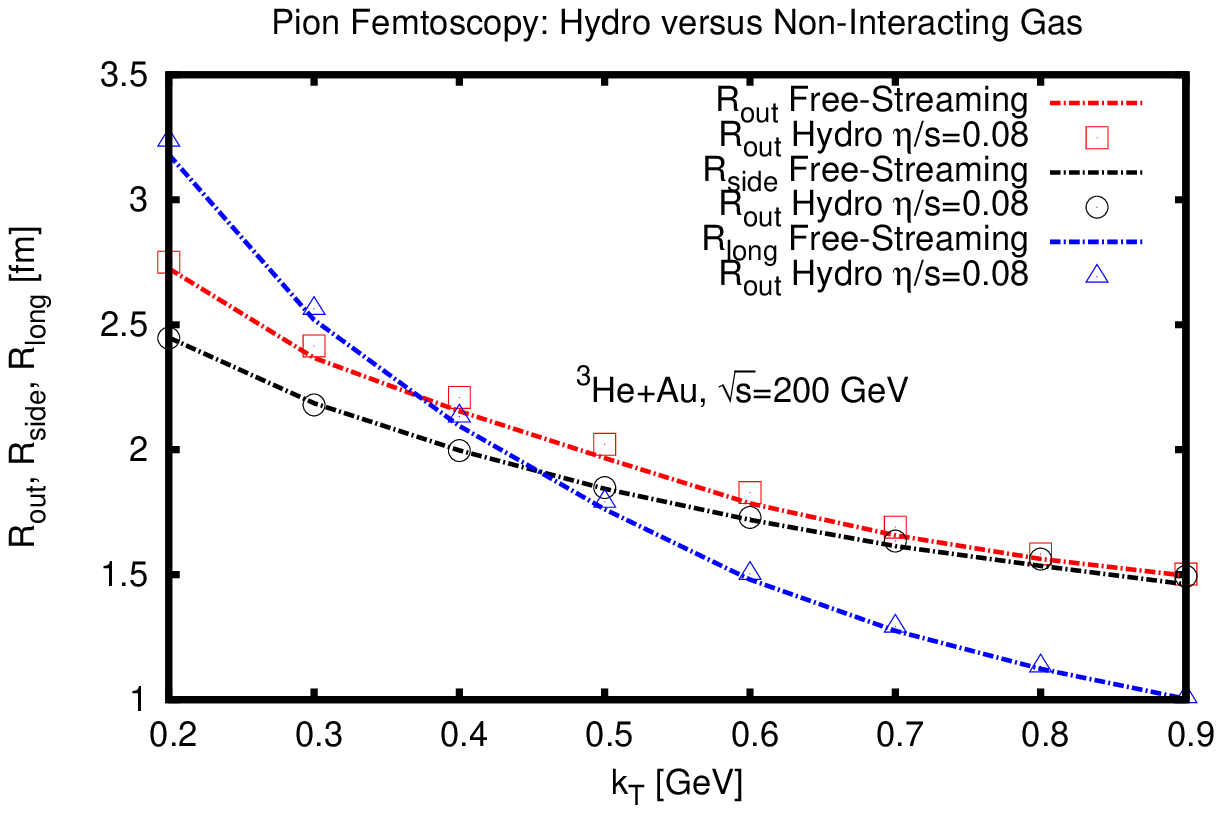}
\hfill
\includegraphics[width=0.45\linewidth]{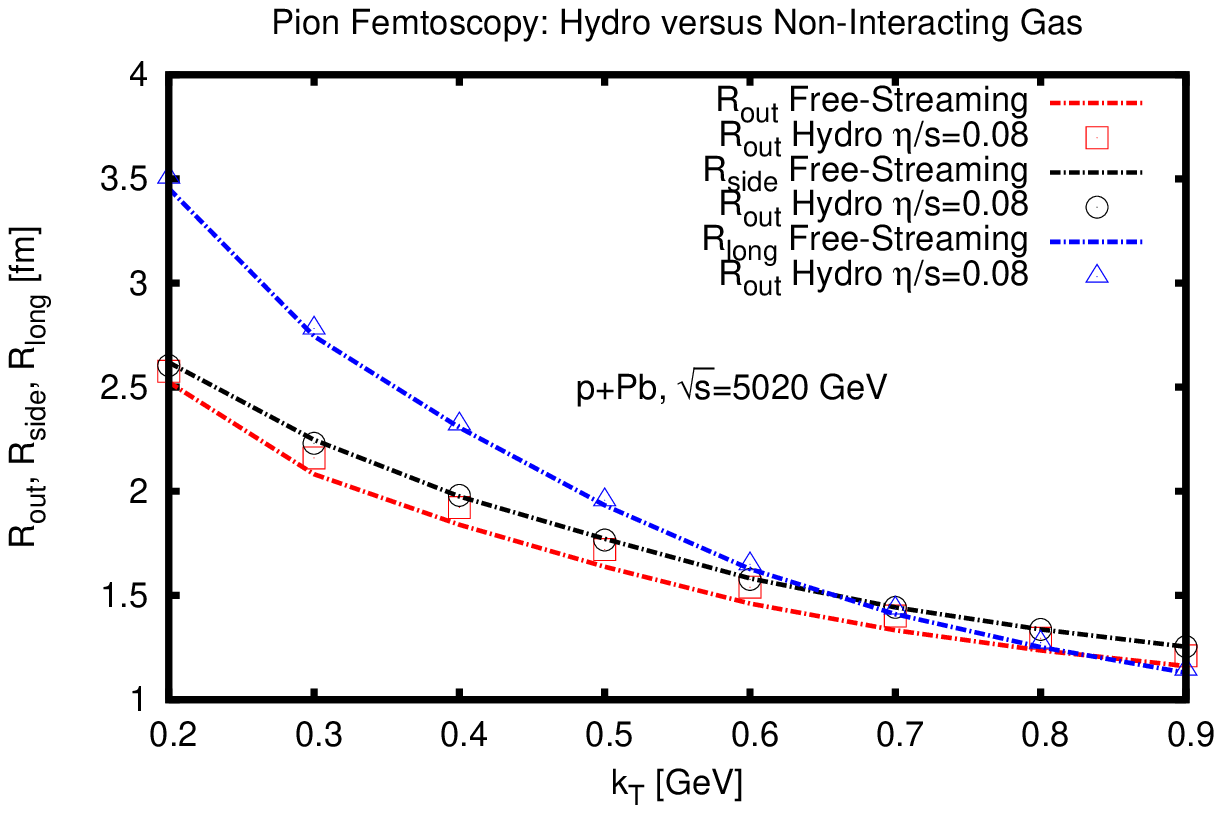}
\hfill
\includegraphics[width=0.45\linewidth]{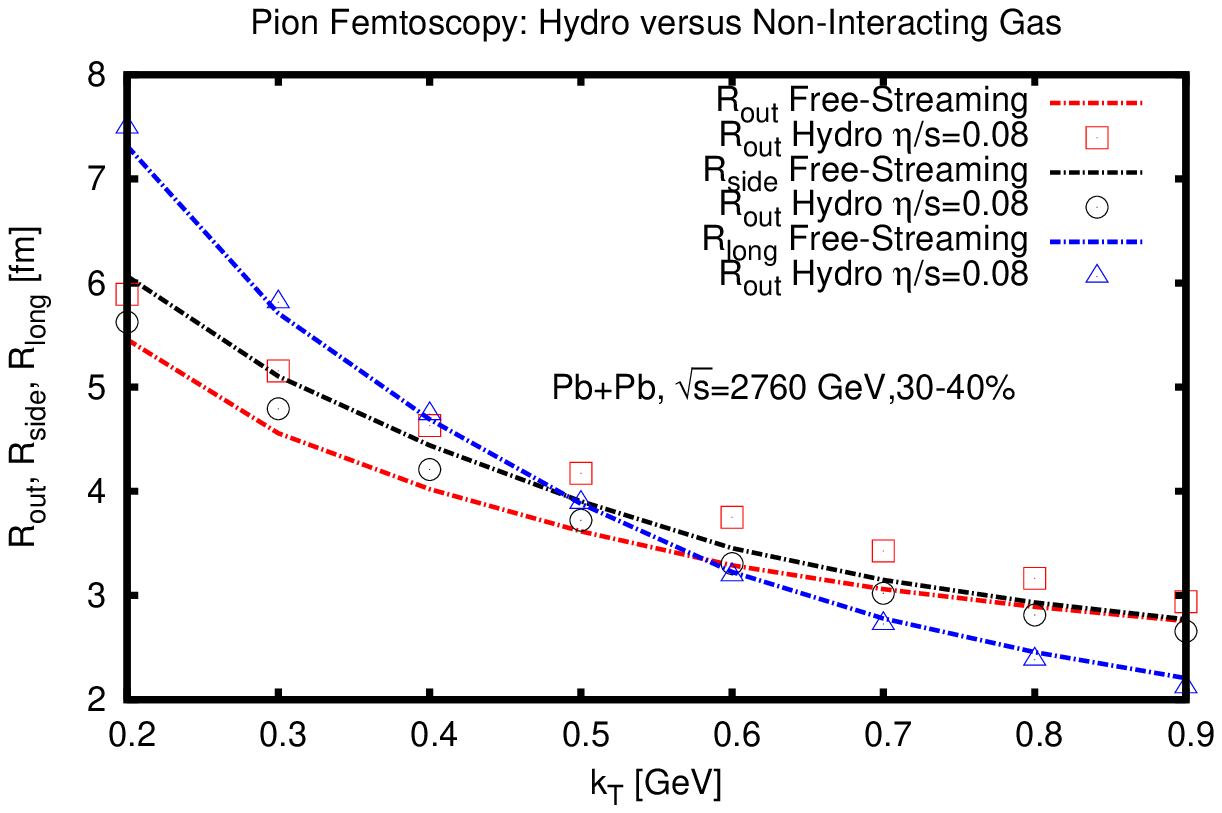}
\caption{\label{fig:hbt} Pion femtoscopic radii (``$R_{\rm out}, R_{\rm side}, R_{\rm long}$'') from simulations of granular \dAu, \hAu, \pPb and \PbPb collisions. Shown are results for identified particles for free-streaming evolution (no-interaction) and almost ideal hydrodynamics ($\eta/s=0.08$), followed by a hadronic cascade.  See text for details. }
\end{figure}

In this work, the femtoscopic measurements are studied through the two-particle correlations \cite{Novak:2013bqa}
\begin{equation}
S({\bf K},{\bf r})\equiv \frac{\int d^3r_1 d^3 r_2 f({\bf K},{\bf r_1}) f({\bf K},{\bf r_2}) \delta\left({\bf r}-({\bf r_1-r_2})\right)}{\int d^3r_1 d^3 r_2 f({\bf K},{\bf r_1}) f({\bf K},{\bf r_2})}\,,
\end{equation}
where $f({\bf K},{\bf r})$ is the particle phase space density in the final state. The information about the correlations is extracted through fitting a Gaussian form to the function $S$, 
\begin{equation}
S({\bf K},{\bf r})\propto e^{-\frac{x^2}{2 R_{\rm out}^2}-\frac{y^2}{2 R_{\rm side}^2}-\frac{z^2}{2 R_{\rm long}^2}}\,
\end{equation}
defining the femtoscopic radii $R_{\rm out},R_{\rm side},R_{\rm long}$. The results for these extracted radii for pions are shown in Fig.~\ref{fig:hbt} for \dAu, \hAu, \pPb and \PbPb collisions, comparing hydrodynamic and non-interacting evolution. From this figure, one can observe a striking similarity for all the extracted radii between strongly interacting evolution (hydrodynamics) and non-interacting evolution (free streaming) for all simulated systems, small and large. Similarly to what was found for the case of radial flow, the femtoscopic radii are essentially insensitive to the details of the system evolution, as long as energy and momentum are conserved.

In essence, this disqualifies the use of pion femtoscopic measurements as serving as evidence for a hydrodynamic phase during the system evolution. 

\subsection{Inverse slope parameters}

In the above result, it was found that radial flow is a feature of both almost ideal hydrodynamics and free-streaming. One might therefore be suspicious that the ubiquitous presence of radial flow in both models indicates a systematic failure of the modeling procedure since it is known from experimental data that radial flow does disappear in 'low' multiplicity p+p and p+A collisions.

\begin{figure}[t]
%\centralizing
\includegraphics[width=0.3\linewidth]{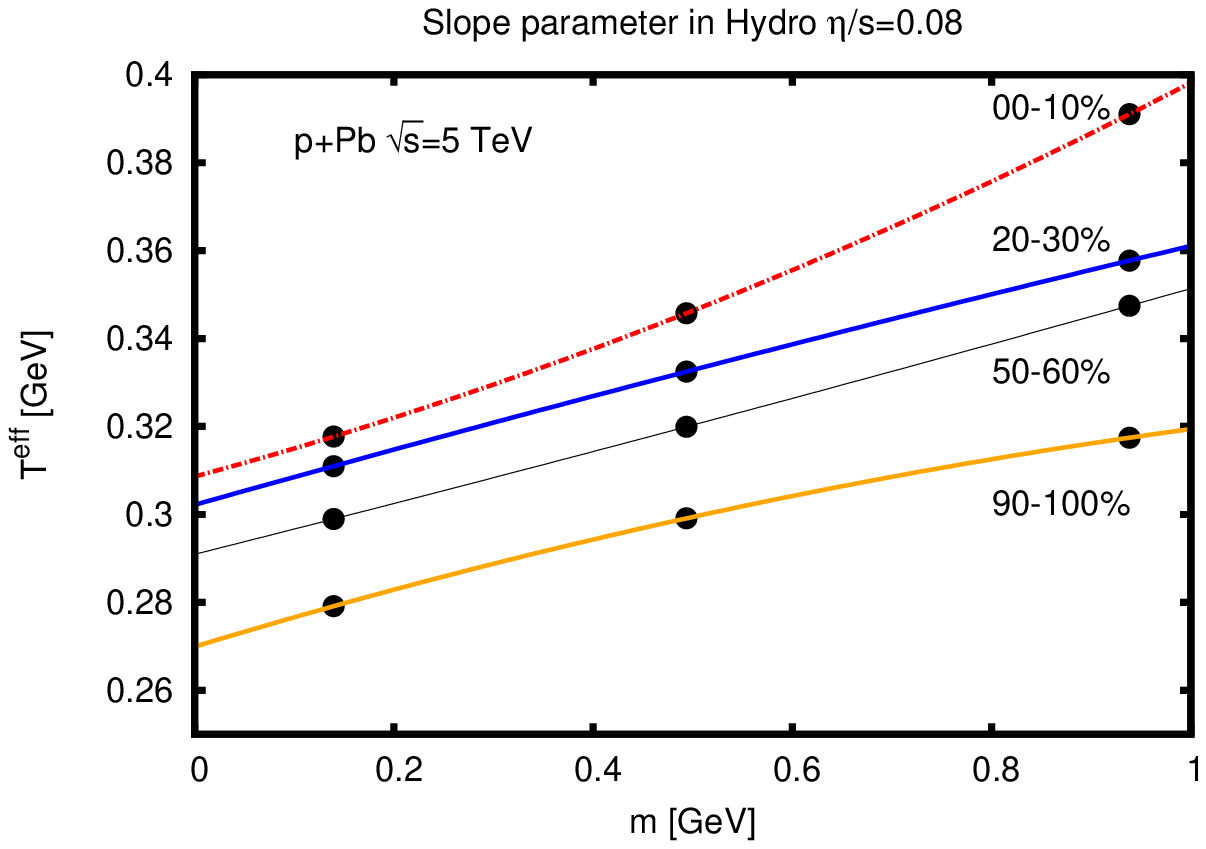}
\hfill
\includegraphics[width=0.3\linewidth]{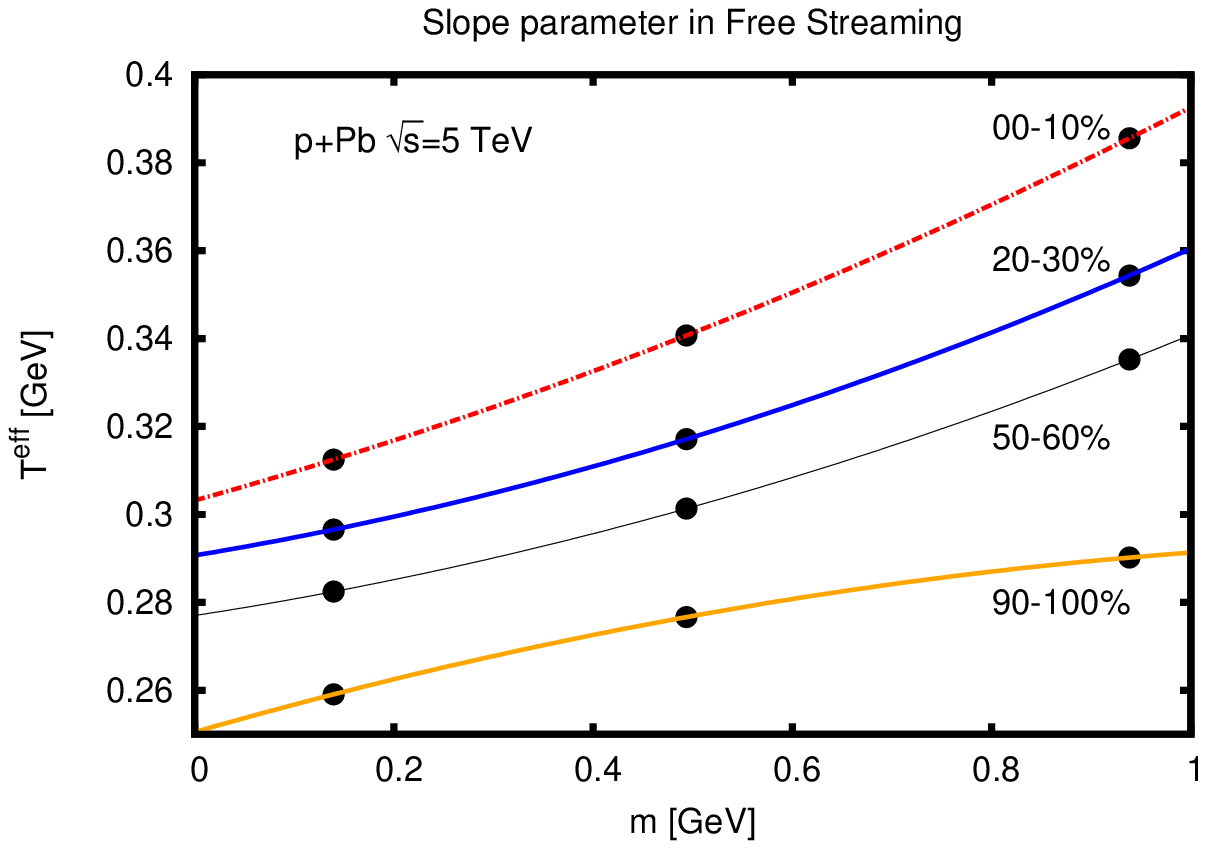}
\hfill
\includegraphics[width=0.2\linewidth]{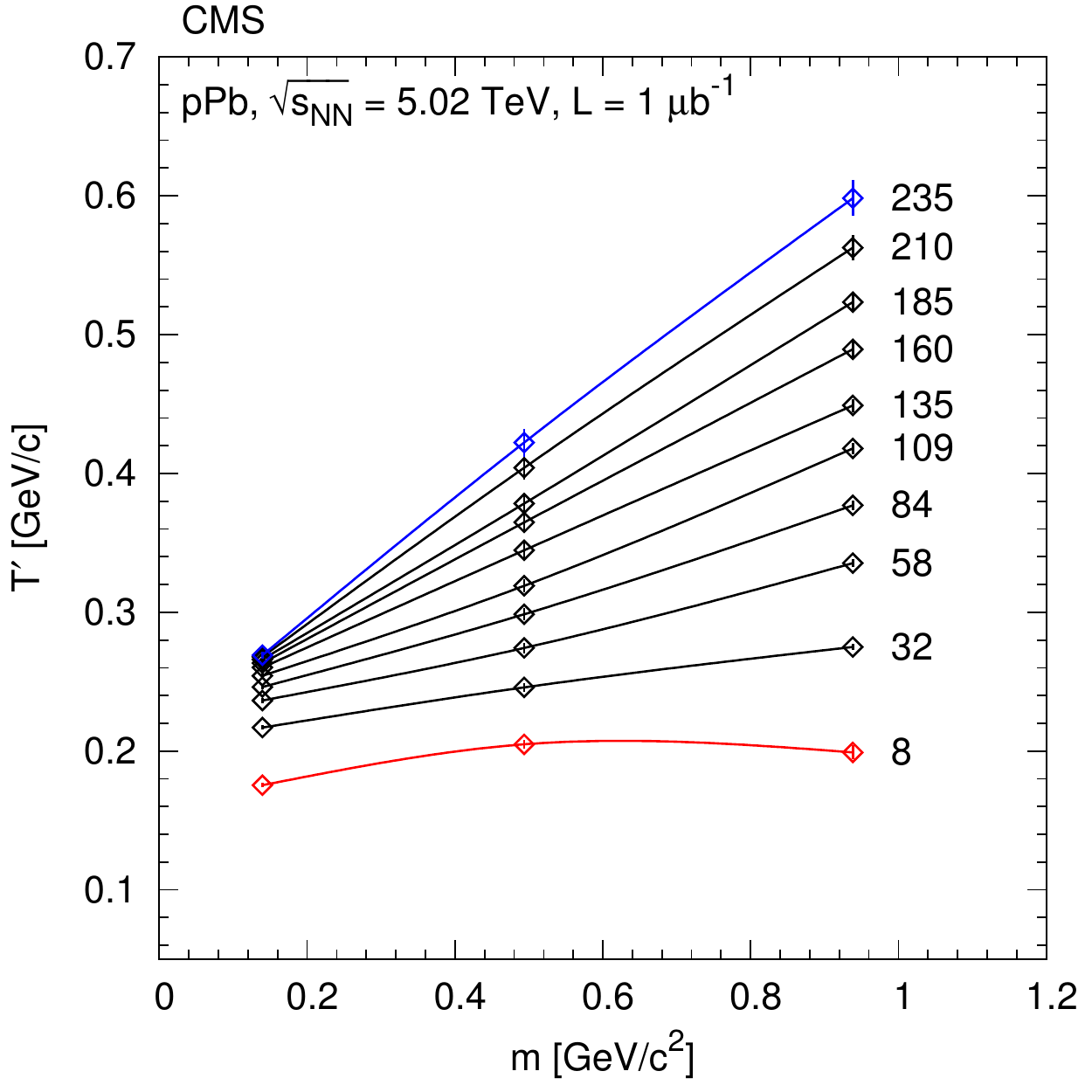}
\hfill
\caption{\label{fig:classify} Effective slope parameter from pion, kaon and proton spectra. Left and middle: Simulation results for almost ideal hydrodynamics and free-streaming for 0-10\%, 20-30\%, 50-60\% and 90-100\% multiplicity bins. Right: Experimental measurement by CMS experiment (figure from Ref.~\cite{Chatrchyan:2013eya}) in terms of total particle tracks (8 to 235).}
\end{figure}

Thus it is interesting to study if radial flow persists in the almost ideal hydrodynamic and free-streaming models if studying low-multiplicity p+A collisions. To this end, \pPb collisions at $\sqrt{s}=5$ TeV were simulated for various multiplicity bins. For each multiplicity bin, the particle spectra for pions, kaons and protons were determined and fit with a form proportional to $\exp\left(-\sqrt{\mu^2+{\bf p}_\perp^2}/T^{\rm eff}\right)$ with $\mu$ the pion, kaon and proton mass, respectively. The effective slope parameter $T^{\rm eff}$ is reported in Fig.\ref{fig:classify} along with the corresponding parameter measured for \pPb collisions at $\sqrt{s}=5$ TeV by the CMS experiment \cite{Chatrchyan:2013eya}. Quantitative results can not be expected to match because the simulation results use an equation of state very different from that of QCD, but qualitative trends should be robust. The experimental data shown in Fig.~\ref{fig:classify} clearly shows that the effective slope parameter dependence on mass decreases significantly from the highest multiplicity selections to the lowest ones. This is often interpreted as a breakdown of collective behavior, as it can be linked to the disappearance of radial flow. 

In the simulation results, especially in the free-streaming model, one can recognize the same qualitative trend: the effective slope parameter dependence on mass decreases for lower multiplicity \pPb events because the system does not 'live' long enough to build up significant amounts of radial flow. Thus the simulation results are not inconsistent with the experimental findings.

However, within the simple geometric Glauber model used for the initial conditions (see section \ref{sec:IS}), it is not possible to realistically describe either the highest multiplicity events (0-1\% or higher) or the lowest multiplicity events (95\%-100\% or lower). In the Glauber model, it is not possible to have less than one collision, which according to section \ref{sec:IS} thus leads to a fixed amount of energy deposition, effectively putting a lower limit of the total multiplicity that can be simulated. Thus, the striking change seen in the experimental data from events with the highest and lowest number of particles cannot be simulated within the simple model for initial conditions adopted here.

\section{Summary and Conclusions}
\label{sec:summ}

In this work flow signatures arising from two very different dynamics in the hot QCD phase following relativistic ion collisions very studied. In the first case, the hot phase dynamics was assumed to be described by non-interacting particles. In the second case, the hot phase dynamics was assumed to be described by extremely strongly interacting modes leading to almost ideal hydrodynamics. In both cases, the exact same initial conditions were implemented and the dynamics was required to correspond to the same equation of state. Also, in both cases the resulting energy-momentum tensor information was recorded on the same space-time grid and then passed on a hadron cascade ``afterburner'' using the same switching procedure. 

%\newpage

Because of this procedure, the resulting particle spectra between the two extreme cases of non-interacting and strongly interacting hot phase dynamics are directly comparable, and lead to the following findings:

\begin{itemize}
\item
Non-interacting (free-streaming) particle dynamics generally leads to equal or larger radial flow than strongly interacting dynamics (hydrodynamics) in all systems considered (\dAu, \hAu, \pPb, \PbPb). Also, as demonstrated in Fig. \ref{fig:two}, radial flow velocities in hydrodynamics and non-interacting dynamics are similar already in the hot QCD phase, suggesting that this result is not dependent on the details of switching or the hadronic cascade. The overall amount of radial flow generated seems to be proportional to the time the systems spend in the hot QCD phase, naturally explaining why radial flow is observed to be very small in e.g. p+p collisions at $\sqrt{s}=200$ GeV (cf. Ref.~\cite{Adams:2003xp}), which have a very short lifetime. {\bf This strongly suggests that the presence of radial flow extracted from experimental measurements should not be used as an indication for the presence of a hydrodynamic phase.}
\item
Non-interacting (free-streaming) particle dynamics generally leads to femtoscopic radii $R_{\rm out},R_{\rm side},R_{\rm long}$ that are very similar to those found in strongly interacting dynamics (hydrodynamics) in all systems considered (\dAu, \hAu, \pPb, \PbPb). {\bf This strongly suggests that the results from femtoscopic measurements should not be used as an indication for the presence of a hydrodynamic phase.}
\item
Non-interacting (free-streaming) particle dynamics generally leads to considerably smaller elliptic flow than strongly interacting dynamics (hydrodynamics) in all systems considered (\dAu, \hAu, \pPb, \PbPb). {\bf This strongly suggests that the presence of a sizable elliptic flow component extracted from experimental measurements is indicative of a hydrodynamic phase.}
\item
Non-interacting (free-streaming) particle dynamics generally leads to triangular and quadrupolar flow components that are comparable or even larger than hydrodynamics in light-on-heavy-ion collisions (\dAu,\hAu and \pPb). This suggests that higher order flow components extracted from experimental measurements for these small systems are not indicative of a hydrodynamic phase during the system evolution. It also suggests that the use of higher order flow components $v_3,v_4,v_5,\ldots$ as a high-precision ``viscometer'' in both small and large systems should be reconsidered because non-hydrodynamic contributions can lead to a considerable contamination of extracted viscosity values. 
\end{itemize}

As an outlook, one should note that generalizations of the free-streaming model description employed here, notably implementations of a QCD equation of state and weak, but non-vanishing interactions are possible should a direct comparison to experimental data become desirable.

\begin{acknowledgments}
 
This work was supported by the Department of Energy, DOE award No. DE-SC0008132. I am thankful for fruitful discussions with Gabriel Denicol, Jamie Nagle, Dhevan Gangadharan, J\"urgen Schuhkraft and Edward Shuryak. This work utilized the Janus supercomputer, which is supported by the National Science Foundation (award number CNS-0821794) and the University of Colorado Boulder. The Janus supercomputer is a joint effort of the University of Colorado Boulder, the University of Colorado Denver and the National Center for Atmospheric Research. Janus is operated by the University of Colorado Boulder.

\end{acknowledgments}

\begin{appendix}
\section{Exact correspondence example between free-streaming and ideal hydrodynamics}
\label{sec:app}

The similarities between non-interacting dynamics (free-streaming) and almost ideal hydrodynamics discussed in the main text may come as a surprise to some readers. However, both dynamics are just different formulations of energy-momentum conservation, so it may not be too surprising to find many similarities between these very different approaches.

To elucidate the power that energy-momentum conservation places on the dynamics, let us give an example that is analytically solvable in both cases: that of SO(3) symmetric flow in Minkowski space (see Ref.~\cite{Bantilan:2012vu} for more details).

To wit, let us study equilibrium initial conditions at $t_0=0$ for a system having a conformal equation of state $T^{\mu}_\mu=\epsilon-3P=0$ with a spherically symmetric initial energy density profile given as
\begin{equation}
\epsilon(t=0,r)=\frac{16 L^4}{(L^2+r^2)^4}\,,\quad L={\rm const.}
\end{equation}
Using spherical coordinates $r,\theta,\phi$ one finds a metric tensor $g_{ab}={\rm diag}(1,-1,-r^2,-r^2\sin^2\theta)$ and the associated non-vanishing Christoffel symbols as
\begin{equation}
\Gamma^{r}_{\theta\theta}=-r\,,\quad
\Gamma^r_{\phi\phi}=-r \sin^2\theta\,,\quad
\Gamma^{\theta}_{r \theta}=\frac{1}{r}\,,\quad
\Gamma^{\theta}_{\phi \phi}=-\cos\theta \sin\theta\,,\quad
\Gamma^\phi_{r \phi}=\frac{1}{r}\,,\quad
\Gamma^{\phi}_{\theta \phi}=\cot \theta\,.
\end{equation}
In these coordinates, the equations for energy momentum conservation are
$\nabla_\mu T^{\mu\nu}=\partial_\mu T^{\mu\nu}+\Gamma^\mu_{\mu \alpha}T^{\alpha \nu}+\Gamma^\nu_{\mu \alpha}T^{\mu \alpha}=0$. Since the problem is spherically symmetric this implies $T^{t\phi}=T^{t \theta}=T^{r\phi}=T^{r\theta}=0$, $T^\theta_\theta=T^\phi_\phi$ and all components of $T^{\mu\nu}$ independent of $\phi$. The equations of motion then simplify to
\begin{eqnarray}
\partial_t \left(r^2 T^{tt}\right)+\partial_r\left(r^2 T^{rt}\right)&=&0\,,\nonumber\\
\partial_t \left(r^3 T^{rt}\right)+\partial_r \left(r^3 T^{rr}\right)-r^2 T^{tt}&=&0\,,\nonumber\\
\partial_\theta T^{\theta \theta}&=&0\,.
\end{eqnarray}
Decomposing the energy-momentum tensor as in Eq.~(\ref{eq:tabhy}), and putting $\Pi=0$ because of symmetry, one finds that a particular solution to the equations of motions and initial conditions is found as \cite{Bantilan:2012vu}
\begin{equation}
\label{hydrosol}
\epsilon(t,r)=\frac{L^4}{\left[t^2 L^2+\frac{(L^2+r^2-t^2)^2}{4}\right]^2}\,,\quad
u^r(t,r)=\frac{r\, t}{\sqrt{t^2 L^2+\frac{(L^2+r^2-t^2)^2}{4}}}\,,\quad \pi^{\mu\nu}=0\,.
\end{equation}
It should be stressed that in this case the shear stress tensor turns out to be zero for all times because the shear velocity gradient is vanishing exactly, so that the present solution is exact for arbitrary values of viscosity.

If one was to calculate spectra for massless particles with Boltzmann statistics at some switching hypersurface $T=T_{SW}$, the hydrodynamic solution would imply 
\begin{equation}
\label{eq:fhydro}
f_{hydro}(t,r,{\bf p})=Z \exp{\left[\frac{-p^0 u^0+p^r u^r}{T_{SW}}\right]}=Z \exp{\left[\frac{-|{\bf p}|\left(L^2+r^2+t^2-2 r\, t \cos\chi\right)}{T_{SW}\sqrt{4 t^2 L^2+(L^2+r^2-t^2)^2}}\right]}\,,
\end{equation}
where $p^r=|{\bf p}|\cos\chi$. The switching hypersurface in this case is located at those points $(t,r)$ where $\epsilon(t,r)={\rm const}$. Choosing the proportionality constant so that $\epsilon=T^4$, we have
\begin{equation}
r^2_{\rm hydro}(t)=t^2-L^2+\frac{2 L \sqrt{1-t^2 T_{SW}^2}}{T_{SW}}\,,\quad T_{SW}<2/L\,.
\end{equation}

Let us now solve the same problem with the same initial conditions for the non-interacting particle case. The initial particle distribution function is given by
\begin{equation}
f_{FS}(t=0,r,{\bf p})=Z \exp{\left[\frac{-|{\bf p}| (L^2+r^2)}{2 L}\right]}\,.
\end{equation}
For Minkowski-space, the free-streaming solution to the Boltzmann equation is easily found to be \hbox{$f(t,r,{\bf p})=f(|{\bf x}-\frac{{\bf p} t}{|{\bf p}|}|)$}, 
so that with above initial condition this leads to
\begin{equation}
f_{FS}(t,r,{\bf p})=Z \exp{\left[\frac{-|{\bf p}| (L^2+r^2+t^2-2 r t \cos\chi)}{2 L}\right]}\,.
\end{equation}
With the choice for $Z$ that is consistent with $\epsilon=T^4$ from above, one finds that the above solution corresponds to an energy density and flow velocity in Eq.~(\ref{hydrosol}). Thus, for a switching hypersurface at $T=T_{SW}$ one finds
\begin{equation}
r_{FS}^2(t)=r_{hydro}^2(t)=t^2-L^2+\frac{2 L \sqrt{1-t^2 T_{SW}^2}}{T_{SW}}\,,\quad T_{SW}<2/L\,.
\end{equation}
At this switching hypersurface, we thus have $2 L = T_{SW}\sqrt{4 t^2 L^2+(L^2+r^2-t^2)^2}$ and thus the free-streaming particle spectrum is
\begin{equation}
\label{eq:equality}
f_{FS}(t,r,{\bf p})=Z \exp{\left[\frac{-p^0 u^0+p^r u^r}{T_{SW}}\right]}=Z \exp{\left[\frac{-|{\bf p}|\left(L^2+r^2+t^2-2 r\, t \cos\chi\right)}{T_{SW}\sqrt{4 t^2 L^2+(L^2+r^2-t^2)^2}}\right]}=f_{\rm hydro}\,.
\end{equation}

Thus, the free-streaming dynamics, including the particle spectrum at the hypersurface, is identical to the ideal hydrodynamic result.

In particular, this proves analytically that free-streaming dynamics (no-interactions, no coupling to hadronic cascades) generates radial flow, as the particle spectra (\ref{eq:equality},\ref{eq:fhydro}) are equal and ideal hydrodynamics does generate radial flow.

\section{Importance of full stress-tensor matching for freeze-out}
\label{app:two}

\begin{figure}[t]
\includegraphics[width=0.7\linewidth]{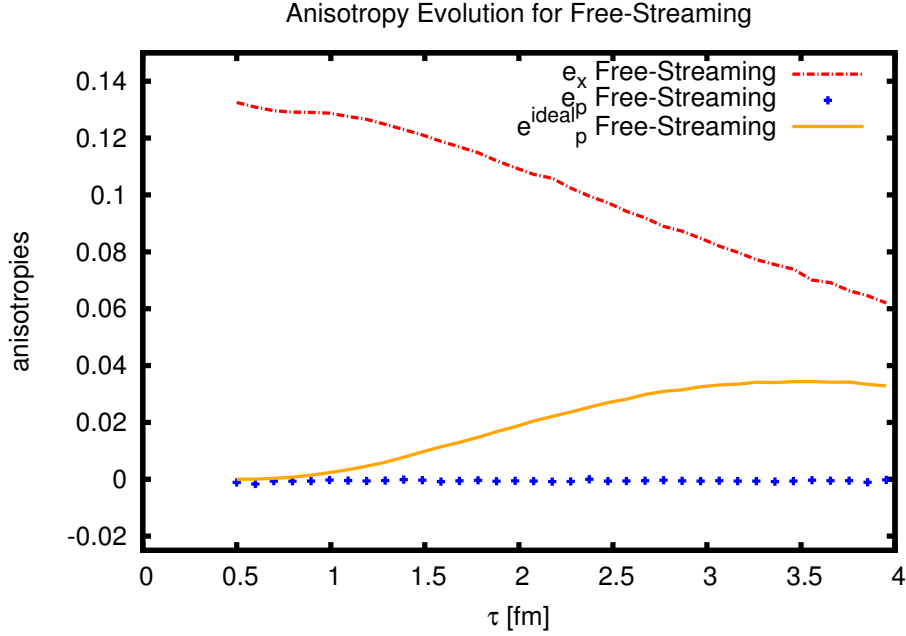}
\caption{\label{fig:last} Time evolution of spatial and momentum anisotropies in free-streaming for smooth initial geometry defined in Sec.\ref{sec:Cu}.}
\end{figure}

As pointed out in the main text, it is important that the full energy-stress tensor is matched when switching from the hot phase to the hadron gas phase. An example of this can be given through monitoring the time-evolution of the momentum anisotropy, defined as
\begin{equation}
e_p\equiv \frac{\int d^2x_\perp T^{xx}-T^{yy}}{\int d^2x_\perp T^{xx}+T^{yy}}\,,
\end{equation}
and its 'ideal hydrodynamic' approximation obtained from dropping all viscous stresses from $T^{ab}$:
\begin{equation}
e_p^{ideal}\equiv \frac{\int d^2x_\perp (\epsilon+P) \left(u^x u^x-u^y u^y\right)}{\int d^2x_\perp(\epsilon+P) \left(u^x u^x+u^y u^y\right)+2P}\,.
\end{equation}
For the case of the smooth collision geometry considered in section \ref{sec:Cu}, the time evolution of these quantities for free-streaming is plotted together in Fig.~\ref{fig:last} with the evolution of the spatial anisotropy
\begin{equation}
e_x\equiv \frac{\int d^2x_\perp \epsilon\left(x^2-y^2\right)}{\int d^2x_\perp \epsilon\left(x^2+y^2\right)}\,.
\end{equation}

From Fig.~\ref{fig:last} one finds that while the full momentum anisotropy $e_p$ is consistent with zero during the entire evolution (as it should be in free-streaming), this is not the case for the ideal hydrodynamic approximation $e_p^{\rm ideal}$. Thus, matching only the ideal hydrodynamic part of the energy-stress tensor in a freeze-out procedure would lead to a 'fake' momentum anisotropy (fake elliptic flow). This has been observed in Ref.~\cite{Broniowski:2008qk}.
By contrast, the matching procedure in Eq.~(\ref{fullmatching}) has been designed to match the full energy-stress tensor, thus preventing fake contribution to arise from the hot phase to hadron gas switching procedure.

Also, the example at hand demonstrates that the net zero momentum anisotropy in free-streaming is a consequence of a delicate cancellation between the (net positive) contribution in the ideal hydrodynamic part $e_p^{\rm ideal}$ and a (net negative) contribution from the viscous stresses.

\end{appendix}

\bibliographystyle{apsrev} \bibliography{allflow}

\end{document}